\documentclass[12pt]{article}

\usepackage{epstopdf,amsfonts,amsmath,amssymb}
\usepackage{mathtools} 
\usepackage{graphicx}
\DeclareGraphicsExtensions{.eps}

\newcommand\encadremath[1]{\vbox{\hrule\hbox{\vrule\kern8pt
\vbox{\kern8pt \hbox{$\displaystyle #1$}\kern8pt}
\kern8pt\vrule}\hrule}}
\def\enca#1{\vbox{\hrule\hbox{
\vrule\kern8pt\vbox{\kern8pt \hbox{$\displaystyle #1$}
\kern8pt} \kern8pt\vrule}\hrule}}

\newcommand\framefig[1]{
\begin{figure}[bth]
\hrule\hbox{\vrule\kern8pt
\vbox{\kern8pt \vbox{
\begin{center}
{#1}
\end{center}
}\kern8pt}
\kern8pt\vrule}\hrule
\end{figure}
}

\newcommand\figureframex[3]{
\begin{figure}[bth]
\hrule\hbox{\vrule\kern8pt
\vbox{\kern8pt \vbox{
\begin{center}
{\mbox{\epsfxsize=#1.truecm\epsfbox{#2}}}
\end{center}
\caption{#3}
}\kern8pt}
\kern8pt\vrule}\hrule
\end{figure}
}
\newcommand\figureframey[3]{
\begin{figure}[bth]
\hrule\hbox{\vrule\kern8pt
\vbox{\kern8pt \vbox{
\begin{center}
{\mbox{\epsfysize=#1.truecm\epsfbox{#2}}}
\end{center}
\caption{#3}
}\kern8pt}
\kern8pt\vrule}\hrule
\end{figure}
}

\makeatletter
\@addtoreset{equation}{section}
\makeatother
\newtheorem{theorem}{Theorem}[section]

\newtheorem{remark}{Remark}[section]
\newtheorem{proposition}{Proposition}[section]
\newtheorem{lemma}{Lemma}[section]
\newtheorem{corollary}{Corollary}[section]
\newtheorem{definition}{Definition}[section]

\newtheorem{example}{Example}[section]
\def\br{\begin{remark}\rm\small}
\def\er{\end{remark}}
\def\bt{\begin{theorem}}
\def\et{\end{theorem}}
\def\bd{\begin{definition}}
\def\ed{\end{definition}}
\def\bp{\begin{proposition}}
\def\ep{\end{proposition}}
\def\bl{\begin{lemma}}
\def\el{\end{lemma}}
\def\bc{\begin{corollary}}
\def\ec{\end{corollary}}
\def\bex{\begin{example}}
\def\eex{\end{example}}

\def\beaq{\begin{eqnarray}}
\def\eeaq{\end{eqnarray}}
\newcommand{\proof}{{\noindent \bf proof:}$\quad$ }
\newcommand{\eproof}{ $\square$ }

\newcommand{\be}{\begin{equation}}
\newcommand{\ee}{\end{equation}}
\newcommand{\beq}{\begin{equation}}
\newcommand{\eeq}{\end{equation}}
\newcommand{\bea}{\begin{eqnarray}}
\newcommand{\eea}{\end{eqnarray}}

\newcommand{\D}{\mathcal{D}}
\newcommand{\diag}{\operatorname{diag}}
\newcommand{\Tr}{{\rm Tr}\,}

\newcommand{\RR}{{\mathbb R}}
\newcommand{\CC}{{\mathbb C}}

\newcommand{\Lieg}{{\mathfrak g}}
\newcommand{\Lieh}{{\mathfrak h}}

\newcommand{\Ad}{\operatorname{Ad}}

\newcommand{\td}{\tilde}
\usepackage[textsize=footnotesize]{todonotes}
\usepackage[pdftex]{hyperref}
\hypersetup{colorlinks,urlcolor=magenta,citecolor=red,linkcolor=blue,filecolor=black}

\newcommand{\examplename}{Example}
\newcommand{\exampleref}[1]{\hyperref[#1]{\examplename~\ref*{#1}}}

\usepackage[style=alphabetic,maxalphanames=4,giveninits,sorting=nyt]{biblatex}
\renewbibmacro{in:}{}
\usepackage{parskip} 
\makeatletter 
\renewcommand\@makefntext[1]{\leftskip=2em\hskip-2em\@makefnmark#1}
\makeatother

\begin{document}

\sloppy

\pagestyle{empty}
\addtolength{\baselineskip}{0.20\baselineskip}
\begin{center}
\vspace{26pt}
{\large \bf {Recursions and ODEs for the correlators in integrable systems and random matrices.}}
\end{center}

\begin{center}

{\sl B.\ Eynard}${}^{1,2}$\hspace*{0.05cm}
{\sl D.\ Mitsios}${}^1$\hspace*{0.05cm}
{\sl S.\ Oukassi}${}^1$\hspace*{0.05cm}

\vspace{6pt}
${}^1$ Universit\'e Paris-Saclay, CEA, CNRS, Institut de physique th\'{e}orique\\91191 Gif-sur-Yvette, France.\\
${}^2$ CRM, Centre de recherches math\'ematiques  de Montr\'eal,\\
Universit\'e de Montr\'eal, QC, Canada.
\end{center}

\vspace{20pt}
\begin{center}
{\bf Abstract}
\end{center}

An integrable system is often formulated as a flat connection, satisfying a Lax equation. It is given in terms of compatible systems having a common solution called the ``wave function" $\Psi$ living in a Lie group $G$, which  satisfies some differential equations with rational coefficients.
From this wave function, it is usual to define a sequence of ``correlators" $W_n$, that play an important role in many applications in mathematical physics.
Here, we show how to systematically obtain ordinary differential equations (ODE) and recursion relations with polynomial coefficients for the correlators.

An application is random matrix theory, where the wave functions are the expectation value of the characteristic polynomial, they form a family of orthogonal polynomials, and are known to satisfy an integrable system. The correlators are then the correlation functions of resolvents or of eigenvalue densities. We give the ODE and recursion on the matrix size that they satisfy. In addition, we discuss generic Fuchsian systems, namely, Schlesinger systems.


%

\vspace{0.5cm}

\vspace{26pt}
\pagestyle{plain}
\setcounter{page}{1}


\tableofcontents
\section{Introduction}
Given a flat section we can describe an integrable system by a Lax equation or equivalently by compatible systems whose common solution is called the ``wave function" $\Psi$ living in a Lie group $G$, this solution satisfies some differential equations with rational coefficients.
Using the wave function we can define a sequence of ``correlators" $(W_n)_{n\in\mathbb{N}}$, that play an important role in many applications in mathematical physics. These correlators were first introduced in the contexte of random matrix theory\cite{BE09}, then used in \cite{BDY1,BDY2} for integrable systems, KDV, KP and intersection numbers. Let's consider a flat section of a reductive Lie group $G$ over $\mathbb{C}P^1$, it is known that the flat section satisfies an equation of the form $\nabla \Psi=0$, taking a faithful representation of the Lie group we can write it as ODE for the wave function, we say in this case that the system is integrable. Furthermore, given the relation between Lie group flat section and adjoint flat section one can derive a Lax equation for this system, then we can define the correlators in their most general form, this allows us to present a general framework of deriving ODEs and recursion relations for the defined correlation functions.

The paper is organized as follows:

\autoref{sec:intro} is devoted to present the ingredients we need for our approach, namely, the flat sections of principal bundle and adjoint bundle, we present the relation between these two flat sections, we give a general definition for the correlators. 

In \autoref{sec:genODE} we present our approach of deriving ODEs for the defined correlators, we start by the correlator $W_1$ then we push our method to higher correlators, in addition to the ODEs, we get that the coefficients are polynomial of all the arguments. 

Then in \autoref{sec:genPDE} we consider a family of systems indexed by a second argument $t$, which gives rise to a system of ODEs for the wave functions, i.e, the family of flat sections of a lie group G principal bundle over $\mathbb C P^1$ with rational connections, we then give the ODEs satisfied by the time depending correlators. 

In \autoref{sec:genRec} we consider this time a family of systems indexed by an integer $N$, we express the relation between different family of correlators $W_n^{(N)}$ by a recursion relation. 

\autoref{sec:1MM} is devoted to applying our formalism to one-matrix model, we start by recalling basic definitions of random matrix theory, we define a matrix wave function via the normalized and dual wave function, we then make the relation between correlators and cumulants of the one-matrix model, from which we derive the ODEs and recursion for these cumulants using the approach developped in the previous sections. 

We recall two-matrix model basic definitions \autoref{sec:2mm} and emphasize its relation with our formalism, we construct linear ODE and recursions for the cumulants of two-matrix model, it is related to the general definition of correlators by a well chosen adjoint flat section. 

In \autoref{sec:MM} we apply our method to particular examples of minimal models, we specialize to the case of $\mathfrak sl(2,\mathbb C)$, starting by the definition of Gelfand-Dikii polynomials, then building Lax pair and applying our approach to derive the ODE with respect to time for correlator $W_1$. 

Schlesinger systems are introduced in \autoref{sec:Ss}, we find the ODE for correlator $W_1$ in the case of $\mathfrak sl(2,\mathbb C)$. 

We gather in \autoref{sec:conc} a number of concluding remarks. 

\section{Preliminaries}
\label{sec:intro}

In short, we consider an invertible matrix $\Psi(x)$ that satisfies a linear differential equation with rational coefficients
$$
\frac{d}{dx}\Psi(x) = \mathcal D(x) \ \Psi(x) \qquad \mathcal D(x)\in M_r(\CC(x)),
$$
where $M_r(\CC(x))$ denotes the set of $r \times r$ matrices with entries rational functions with complex coefficients.
Invertibility of $\Psi(x)$ suggests that it belongs to a representation of the Lie group $GL(r,\CC)$ for each $x\in \CC $. Therefore, it is locally a holomorphic section of a bundle over $\CC P^1 = \CC \cup \{\infty\}$, whose fiber is the Lie group $G=GL(r,\CC)$. $\nabla = d-\mathcal D(x)dx$ is a rational connexion on this bundle and $\Psi(x)$ is a flat section.

\subsection{Differential systems and flat sections of Lie groups}

More generally, we consider  $\Psi(x)$ a flat section (we mean a local section that can be analytically continued to be globally multivalued) of a reductive Lie group $G$ trivial\footnote{In this article we restrict ourselves to trivial $G$-bundles over $\CC P^1$. All the formalism can be extended to more general cases but requires more background, and we leave it for future work. The case studied here has already a large number of practical applications.} holomorphic bundle over $\CC P^1$, with a meromorphic connection:
\beq
\nabla \Psi = 0, \quad
\nabla = d-\D(x) dx,
\eeq
where $\D(x)$ is valued in the adjoint bundle, which implies that for each $x$,  $\D(x)$ is in the Lie algebra $\Lieg$ of the group $G$, and its entries are meromorphic functions of $x$ (i.e. rational) with possibly some poles, i.e.,
\beq
\D(x) \in \Lieg \otimes \CC(x).
\eeq
$\Psi(x)$ being a flat section can be written in some faithful matrix representation of $G$, as:
\beq
\frac{d}{dx}\Psi(x) = \D(x) \Psi(x) \qquad \, \ \ \Psi(x)\in G.
\eeq

\bex[$\Psi \in GL(r,\CC)$]
Consider the case $G=GL(r,\CC)$, then $\Psi(x)$ is an invertible $r\times r$ matrix,  and $\Lieg=\mathfrak gl(r,\CC)=M_r(\CC) $,  and $\D(x)$ is an $r\times r$ matrix, whose elements are rational functions of $x$ (polynomial if the only pole is at $\infty$).
\eex

\bex[Airy differential system]
Choose $G=SL(2,\CC)$ ($2\times 2$ matrices of determinant 1), with $\Lieg=\mathfrak sl(2,\CC)$ ($2\times 2$ matrices of trace 0), and
\beq
\D(x) = \begin{pmatrix}
0 & 1 \cr x & 0
\end{pmatrix}.
\eeq
A flat section $\Psi(x)$ is solution of
\beq
\nabla \Psi = 0
\quad \Leftrightarrow \quad
\frac{d}{dx} \Psi(x) = \D(x) \Psi(x),
\eeq
it is expressed in terms of two linearly independent solutions of the Airy differential equation
$
\big(\frac{d^2}{dx^2} - x \big)f(x) = 0
$
called the Airy and Bairy functions, denoted as $Ai(x)$ and $Bi(x)$ respectively.
\beq
\Psi(x) = \begin{pmatrix}
Ai(x) & Bi(x) \cr Ai'(x) & Bi'(x)
\end{pmatrix}
\in G=SL(2,\CC).
\eeq
Additionally, the Wronskian of $Ai(x)$ and $Bi(x)$ is equal to $1$, namely, 
\beq
\det\Psi(x) = Ai(x)Bi'(x)-Ai'(x)Bi(x) = 1.
\eeq

\eex

\bex[Hermite differential system, also called GUE]
Let $N>0$ be an integer.
In the case where $G=GL(2,\CC)$, with $\Lieg=\mathfrak gl(2,\CC)$, and
\beq
\D(x) = \begin{pmatrix}
x & -1 \cr N & 0
\end{pmatrix}.
\eeq
The flat section $\Psi(x)$ is a solution of
\beq
\nabla \Psi = 0
\quad \Leftrightarrow \quad
\frac{d}{dx} \Psi(x) = \D(x) \Psi(x),
\eeq
given by
\beq
\Psi(x) = \begin{pmatrix}
H_{N-1}(x) & \td H_{N-1}(x) \cr H_N(x) & \td H_N(x)
\end{pmatrix},
\eeq
where $H_N(x)$ is the $N^{\rm th}$ Hermite polynomial, and $\td H_N$ is the $N^{th}$ Hermite function (which is not a polynomial of $x$).
Both $H_N$ and $\td H_N$ are solutions of the equation:
\beq
f''(x)-x f'(x)+Nf(x)=0.
\eeq
\eex
%
%

\subsection{Adjoint Flat sections}

The connection $\nabla$ acts in the adjoint bundle by the commutator, we define $\nabla_{\text{adj}}=d-[\D dx,.]$.
A flat section $M(x)$ in the adjoint bundle ($M(x)\in \Lieg$ for each $x$), must satisfy:
\beq
\nabla_{\text{adj}} M = 0,
\eeq
in other words
\beq
\frac{d}{dx} M(x) =[\D(x),M(x)].
\eeq

\bl
Let $\Psi(x)$ be a flat section of the group bundle. Every adjoint flat section $M(x)$ must be of the form:
\beq
M(x) = \Psi(x) E \Psi(x)^{-1}  = \Ad_{\Psi(x)} E,
\eeq
where $E\in\Lieg$ is constant $dE/dx=0$.

\el
\proof{}This is a classical result.
Let us redo it here, in a matrix representation:
\beq
\begin{aligned}
\frac{d}{dx} M(x)
&=&& \frac{d}{dx} \left( \Psi(x) E \Psi(x)^{-1} \right) \cr
&=&& \frac{d}{dx} \Psi(x) \ \  E \Psi(x)^{-1} + \Psi(x) \ \  E \frac{d}{dx} \Psi(x)^{-1}  + \Psi(x) \frac{dE}{dx}  \Psi(x)^{-1}\cr
&=&& \frac{d}{dx} \Psi(x) \ \  E \Psi(x)^{-1} - \Psi(x) \ \  E \Psi(x)^{-1} \frac{d}{dx}\Psi(x) \ \Psi(x)^{-1}  + \Psi(x) \frac{dE}{dx}  \Psi(x)^{-1}\cr
&=&& \left[\frac{d}{dx} \Psi(x) \Psi(x)^{-1} ,\Psi(x) \ \  E \Psi(x)^{-1} \right]  + \Psi(x) \frac{dE}{dx}  \Psi(x)^{-1}\cr
&=&& \left[\mathcal D(x) ,\ M(x) \right]  + \Psi(x) \frac{dE}{dx}  \Psi(x)^{-1}.
\end{aligned}
\eeq
It satisfies $dM/dx=[\mathcal D,M]$ if and only if $dE/dx=0$.
\eproof

\bd[$M(x.E)$ Adjoint flat sections]\label{def:AdjointFlatSection}
For each $E\in\Lieg $, we shall denote this section as $M(x.E)$:
\beq
M(x.E) = \Ad_{\Psi(x)}E.
\eeq
We can also write it:
\beq
M(X),
\eeq
where $X=x.E\in \CC P^1\times \Lieg $ is a pair of a point and a Lie algebra element, i.e. a Lie algebra weighted point.

\ed
The intuitive way of thinking of it, is that $E$ is a vector in the Lie algebra $\Lieg$, more precisely the $\Lieg$-fiber of the adjoint bundle at the ``base point" used to define the universal cover of $\CC P^1\setminus\text{poles}$, and then $M(x.E)$ is the parallel transport of $E$ from the base point to the fiber over the point $x$.

\subsection{Definition of Correlators}

We define the following functions, as they are very useful in many applications of these frameworks, for example in random matrix theory or many integrable systems.
They were first introduced in \cite{BE09} and subsequent works, and popularized by \cite{BDY1,BDY2}.

For this, we choose a faithful matricial representation of the group $G$ into  $GL(r,\CC)$ (invertible $r\times r$ matrices), and $\Lieg$ represented as $r\times r$ matrices in $
\Lieg l(r,\CC)$.
We consider the Trace $\Tr$ in this representation.

\bd[Correlators]\label{def:Wngeneral}
Let $\Psi(x)$ be a once for all chosen flat section of the group bundle.
Let $M(x.E)=\Psi(x)E\Psi(x)^{-1}$ be a flat section of the adjoint bundle, parametrized by  $E\in \Lieg$.
We define:
\beq
W_1(x.E) = \Tr \D(x) M(x.E),
\eeq
\beq
W_2(x_1.E_1,x_2.E_2) = \frac{1}{(x_1-x_2)^2} \Tr M(x_1.E_1)M(x_2.E_2),
\eeq
and for $n\geq 3$
\beq\label{eq:defWn}
W_n(x_1.E_1,\dots,x_n.E_n) = \sum_{\sigma\in \mathfrak S_n^{1-\text{cycle}}}
(-1)^\sigma \frac{\Tr \prod_{i=1}^n M(x_{\sigma^i(1)}.E_{\sigma^i(1)})}{\prod_{i=1}^n (x_{\sigma(i)}-x_i)},
\eeq
where $\mathfrak S_n^{1-\text{cycle}}$ is the subset of $\mathfrak S_n$ of permutations having just 1-cycle.

\ed

\bex[Correlator $W_3$]
Recalling that $X_i=x_i.E_i$ we have
\beq
\begin{aligned}
W_3(X_1,X_2,X_3)
&=&& \frac{\Tr M(X_1)M(X_2)M(X_3)}{(x_2-x_1)(x_3-x_2)(x_1-x_3)}
+\frac{\Tr M(X_1)M(X_3)M(X_2)}{(x_3-x_1)(x_2-x_3)(x_1-x_2)} \cr
&=&& \frac{\Tr M(X_1) [M(X_2),M(X_3)]}{(x_2-x_1)(x_3-x_2)(x_1-x_3)}
\end{aligned}
\eeq
and so on for higher $n$.
\eex
Our goal now is to exhibit the differential equations satisfied by these $W_n$s.
\br
We will use in several occasions the fact that $\mathfrak{g}\otimes\CC(x)$ is of dimension $dim \mathfrak{g}$, since the field extension of the finite dimensional Lie algebra is of the same dimension.
\er

\section{ODEs for Correlators}
\label{sec:genODE}

\subsection{ODE for \texorpdfstring{$W_1$}{W1}}

Let us warm up with $W_1(X)$ with $X=x.E$. Let $'$ denote $d/dx$.
We have
\beq
\begin{aligned}
W'_1(X)
&=&& \Tr \D'(x) M(X) + \Tr \D(x) M'(X) \cr
&=&& \Tr \D'(x) M(X) + \Tr \D(x) [\D(x),M(X)] \cr
&=&& \Tr \D'(x) M(X) + \Tr [\D(x), \D(x)] M(X) \qquad \cr
&=&& \Tr \D'(x) M(X), 
\end{aligned}
\eeq
where the previous to last equality follows due to cyclicity of the trace. \\
Then
\beq
\begin{aligned}
W''_1(X)
&=&& \Tr \D''(x) M(X) + \Tr \D'(x) M'(X) \cr
&=&& \Tr \D''(x) M(X) + \Tr \D'(x) [\D(x),M(X)] \cr
&=&& \Tr \D''(x) M(X) + \Tr [\D'(x), \D(x)] M(X) \qquad \cr
&=&& \Tr \left(\D''(x) + [\D'(x),\D(x)]\right) M(X) .
\end{aligned}
\eeq
We can repeat this and get
\beq
\frac{d^k}{dx^k} W_1(X) = \Tr \D_k(x) M(X),
\eeq
with the matrix $\D_k(x)$ computed as follows
\beq
\D_0=\D, \quad  \D_1 = \D', \quad  \D_2 = \D'' + [\D',\D], \quad  \D_3 = \D''' + 2 [\D'',\D] + [[\D',\D],\D]
\eeq
and recursively
\beq
\D_{k+1} = \D'_k + [\D_k,\D].
\eeq
Now notice that each $\D_k(x)$ is a rational function of $x$ valued in the Lie algebra $\Lieg$, that is a finite dimensional vector space of dimension $\dim \Lieg$.
Therefore, at most $\dim\Lieg$ of them can be linearly independent over the field $\CC(x)$
which implies that there must exist some rational coefficients $\alpha_k(x)$, such that
\beq
\sum_{k=0}^{\dim\Lieg} \alpha_k(x) \D_k(x)=0.
\eeq
And upon multiplying by the common denominator, we can assume that $\alpha_k(x)\in \CC[x]$ are polynomial.
Therefore,
\bt \label{thm:ODEforW1}
$W_1(x)$ satisfies a linear differential equation of order $\dim\Lieg$, with polynomial coefficients $\alpha_k(x)$:
\beq
\sum_{k=0}^{\dim\Lieg}  \alpha_k(x) \frac{d^k}{dx^k} W_1(x.E) = 0.
\eeq
The coefficients $\alpha_k(x)$ are determined by
\beq
\sum_{k=0}^{\dim\Lieg}  \alpha_k(x) \D_k(x) = 0
\qquad \text{where} \quad
\D_0 = \D, \quad \D_{k+1}=\D_k' + [\D_k,\D].
\eeq
\et

Notice that the matrices $\D_k(x)$ are independent of $E$, and so are the $\alpha_k(x)$.
This implies that for any choice of $E\in \Lieg$ we have a solution of this differential equation.
Therefore,
\bt
The space of solutions of
\beq
\left(\sum_{k=0}^{\dim\Lieg}  \alpha_k(x) \frac{d^k}{dx^k} \right) w(x) = 0
\eeq
is the vector space of dimension $\dim \Lieg$ generated by
\beq
W_1(x.E) = \Tr \D(x) M(x.E)
\qquad \text{for } E\in \Lieg.
\eeq

\et

\bex[Airy]
We have $\Lieg=\mathfrak sl(2)$ which has dimension $\dim \mathfrak sl(2)=3$:
\beq
\D_0(x) = \begin{pmatrix}
0 & 1 \cr x & 0
\end{pmatrix}, \ \
\D_1(x) = \begin{pmatrix}
0 & 0 \cr 1 & 0
\end{pmatrix}, \ \
\D_2(x) = \begin{pmatrix}
-1 & 0 \cr 0 & 1
\end{pmatrix}, \ \
\D_3(x) = \begin{pmatrix}
0 & -2 \cr 2x & 0
\end{pmatrix}.
\eeq
They satisfy the linear relation
\beq
\frac12 \D_3 = 2x \D_1 - \D_0,
\eeq
by the use of \autoref{thm:ODEforW1}
\beq
\frac12 W_1''' = 2x W_1' - W_1.
\eeq
The general solution to this equation is:
\beq
W_1(x.E),
\eeq
with $E$ any linear combination of the three Chevaley--basis\footnote{In fact any basis of $\Lieg$ would be good. The Chevaley basis is a cannonical basis defined for any semi-simple Lie algebra.} vectors:
\beq
\begin{pmatrix}
0 & 1 \cr 0 & 0
\end{pmatrix}, \quad
\begin{pmatrix}
0 & 0 \cr 1 & 0
\end{pmatrix}, \quad
\begin{pmatrix}
\frac12 & 0 \cr 0 & -\frac12
\end{pmatrix}.
\eeq

\eex

\bex[Hermite/GUE]
We have $\Lieg=\mathfrak gl(2)$ with dimension $\dim \mathfrak gl(2)=4$:
\beq
\D_0(x) = \begin{pmatrix}
x & -1 \cr N & 0
\end{pmatrix}, \ \
\D_1(x) = \begin{pmatrix}
1 & 0 \cr 0 & 0
\end{pmatrix}, \ \
\D_2(x) = \begin{pmatrix}
0 & -1 \cr -N & 0
\end{pmatrix},
\eeq
\beq
\D_3(x) = \begin{pmatrix}
-2N & x \cr -Nx & 2N
\end{pmatrix}, \ \
\D_4(x) = \begin{pmatrix}
0 & 4N-x^2+1 \cr 4N^2-Nx^2-N & 0
\end{pmatrix}.
\eeq
The linear relation is
\beq
\D_4 = (x^2-4 N) \D_2 + x \D_1 -\D,
\eeq
using \autoref{thm:ODEforW1}
\beq
W_1'''' = (x^2-4N) W_1''+xW_1' - W_1 .
\eeq
We shall come back to this equation and its consequences in \exampleref{ex:GUE1MM}.
\eex

\subsection{ODE for \texorpdfstring{$W_2$}{W2}}

Let $X_1=x_1.E_1$ and $X_2=x_2.E_2$.
We have defined
\beq
W_2(X_1,X_2) =  \Tr \frac{M(X_2)}{(x_1-x_2)^2} \ M(X_1).
\eeq
Taking the derivative with respect to $x_1$ we get
\beq
\frac{d}{dx_1} W_2(X_1,X_2) =  \Tr \left(\frac{[M(X_2),\D(x_1)]}{(x_1-x_2)^2}- \frac{2M(X_2)}{(x_1-x_2)^3} \right) \ M(X_1).
\eeq

And in general
\beq
\frac{d^k}{dx_1^k} W_2(X_1,X_2) =  \Tr \D_k(x_1;X_2) \ M(X_1),
\eeq
where
\beq
\D_0(x_1;X_2) = \frac{M(X_2)}{(x_1-x_2)^2}, \quad
\D_{k+1}(x_1;X_2) = \frac{d}{dx_1}\D_k(x_1;X_2) + [ \D_k(x_1;X_2),\D(x_1)].
\eeq
Remark that for each $k$, $\D_k(x_1;X_2)$ is a rational function of $x_1$, but it may be a transcendantal function of $x_2$ (in the Airy example it involves $Ai(x_2)$).
Again, no more than $\dim\Lieg$ of them can be linearly independent, and therefore there exist some coefficients $\alpha_k(x_1;X_2)$ polynomial in $x_1$, such that
\beq
\sum_{k=0}^{\dim \Lieg} \alpha_k(x_1;X_2) \D_k(x_1;X_2) = 0.
\eeq
This implies,
\bt
$W_2(X_1,X_2)$ satisfies an $x_1$--differential equation of order $\dim\Lieg$, with  coefficients $\alpha_k(x_1;X_2)$ polynomials of $x_1$:
\beq
\sum_{k=0}^{\dim\Lieg}  \alpha_k(x_1;X_2) \frac{d^k}{dx_1^k} W_2(X_1,X_2) = 0.
\eeq

\et

\bex[Airy]
Using numerical computation we find the following expression for $\alpha_k$s in the case of Airy, the coefficients depend on the choice of $E_2$, in our case we choose as an example:
\beq
E_2=\begin{pmatrix}
0&1\\
0&0
\end{pmatrix},
\eeq 
and the coefficients are:

\begin{equation}
\begin{aligned}
\alpha_0 (x_1,x_2)
&=
- 6  x_{1}^{4} \operatorname{Ai}^{4}{\left(x_{2} \right)}
+ 4  x_{1}^{3} x_{2} \operatorname{Ai}^{4}{\left(x_{2} \right)}
+ 16  x_{1}^{3} \operatorname{Ai}^{2}{\left(x_{2} \right)} {\operatorname{Ai}^{'}}^{2}{\left(x_{2} \right)}
\\ & \quad
+ 2  x_{1}^{2} x_{2}^{2} \operatorname{Ai}^{4}{\left(x_{2} \right)}
- 16  x_{1}^{2} x_{2} \operatorname{Ai}^{2}{\left(x_{2} \right)} {\operatorname{Ai}^{'}}^{2}{\left(x_{2} \right)}
- 2  x_{1}^{2} \operatorname{Ai}^{3}{\left(x_{2} \right)} \operatorname{Ai}^{'}{\left(x_{2} \right)}
\\ & \quad
- 10  x_{1}^{2} {\operatorname{Ai}^{'}}^{4}{\left(x_{2} \right)}
- 4  x_{1} x_{2} \operatorname{Ai}^{3}{\left(x_{2} \right)} \operatorname{Ai}^{'}{\left(x_{2} \right)}
+ 12  x_{1} x_{2} {\operatorname{Ai}^{'}}^{4}{\left(x_{2} \right)}
\\ & \quad
+ 8  x_{1} \operatorname{Ai}{\left(x_{2} \right)} {\operatorname{Ai}^{'}}^{3}{\left(x_{2} \right)}
+ 6  x_{2}^{2} \operatorname{Ai}^{3}{\left(x_{2} \right)} \operatorname{Ai}^{'}{\left(x_{2} \right)}
- 2  x_{2}^{2} {\operatorname{Ai}^{'}}^{4}{\left(x_{2} \right)}
\\ & \quad
- 6  x_{2} \operatorname{Ai}^{4}{\left(x_{2} \right)}
- 8  x_{2} \operatorname{Ai}{\left(x_{2} \right)} {\operatorname{Ai}^{'}}^{3}{\left(x_{2} \right)}
+ 6  \operatorname{Ai}^{2}{\left(x_{2} \right)} {\operatorname{Ai}^{'}}^2{\left(x_{2} \right)}
\\
\alpha_1(x_1,x_2)
&=
- 4  x_{1}^{5} \operatorname{Ai}^{4}{\left(x_{2} \right)}
+ 8  x_{1}^{4} x_{2} \operatorname{Ai}^{4}{\left(x_{2} \right)}
+ 8  x_{1}^{4} \operatorname{Ai}^{2}{\left(x_{2} \right)} {\operatorname{Ai}^{'}}^{2}{\left(x_{2} \right)}
\\ & \quad
- 4  x_{1}^{3} x_{2}^{2} \operatorname{Ai}^{4}{\left(x_{2} \right)}
- 16  x_{1}^{3} x_{2} \operatorname{Ai}^{2}{\left(x_{2} \right)} {\operatorname{Ai}^{'}}^{2}{\left(x_{2} \right)}
- 4  x_{1}^{3} \operatorname{Ai}^{3}{\left(x_{2} \right)} \operatorname{Ai}^{'}{\left(x_{2} \right)}
\\ & \quad
- 4  x_{1}^{3} {\operatorname{Ai}^{'}}^{4}{\left(x_{2} \right)}
+ 8  x_{1}^{2} x_{2}^{2} \operatorname{Ai}^{2}{\left(x_{2} \right)} {\operatorname{Ai}^{'}}^{2}{\left(x_{2} \right)}
+ 8  x_{1}^{2} x_{2} \operatorname{Ai}^{3}{\left(x_{2} \right)} \operatorname{Ai}^{'}{\left(x_{2} \right)}
\\ & \quad
+ 8  x_{1}^{2} x_{2} {\operatorname{Ai}^{'}}^{4}{\left(x_{2} \right)}
- 3  x_{1}^{2} \operatorname{Ai}^{4}{\left(x_{2} \right)}
+ 4  x_{1}^{2} \operatorname{Ai}{\left(x_{2} \right)} {\operatorname{Ai}^{'}}^{3}{\left(x_{2} \right)}
\\ & \quad
- 4  x_{1} x_{2}^{2} \operatorname{Ai}^{3}{\left(x_{2} \right)} \operatorname{Ai}^{'}{\left(x_{2} \right)}
- 4  x_{1} x_{2}^{2} {\operatorname{Ai}^{'}}^{4}{\left(x_{2} \right)}
+ 6  x_{1} x_{2} \operatorname{Ai}^{4}{\left(x_{2} \right)}
\\ & \quad
- 8  x_{1} x_{2} \operatorname{Ai}{\left(x_{2} \right)} {\operatorname{Ai}^{'}}^{3}{\left(x_{2} \right)}
+ 3  x_{2}^{2} \operatorname{Ai}^{4}{\left(x_{2} \right)}
+ 4  x_{2}^{2} \operatorname{Ai}{\left(x_{2} \right)} {\operatorname{Ai}^{'}}^{3}{\left(x_{2} \right)}
\\ & \quad
- 12  x_{2} \operatorname{Ai}^{2}{\left(x_{2} \right)} {\operatorname{Ai}^{'}}^{2}{\left(x_{2} \right)}
+ 6  {\operatorname{Ai}^{'}}^4{\left(x_{2} \right)}
\\
\alpha_2(x_1,x_2)
&=
3  x_{1}^{3} \operatorname{Ai}^{4}{\left(x_{2} \right)}
- 9  x_{1}^{2} \operatorname{Ai}^{2}{\left(x_{2} \right)} {\operatorname{Ai}^{'}}^{2}{\left(x_{2} \right)}
- 3  x_{1} x_{2}^{2} \operatorname{Ai}^{4}{\left(x_{2} \right)}
\\ & \quad
+ 6  x_{1} x_{2} \operatorname{Ai}^{2}{\left(x_{2} \right)} {\operatorname{Ai}^{'}}^{2}{\left(x_{2} \right)}
+ 6  x_{1} {\operatorname{Ai}^{'}}^{4}{\left(x_{2} \right)}
+ 3  x_{2}^{2} \operatorname{Ai}^{2}{\left(x_{2} \right)} {\operatorname{Ai}^{'}}^{2}{\left(x_{2} \right)}
\\ & \quad
- 6  x_{2} {\operatorname{Ai}^{'}}^4{\left(x_{2} \right)}\\
\alpha_3(x_1,x_2)
&=
x_{1}^{4} \operatorname{Ai}^{4}{\left(x_{2} \right)}
- 2 x_{1}^{3} x_{2} \operatorname{Ai}^{4}{\left(x_{2} \right)}
- 2 x_{1}^{3} \operatorname{Ai}^{2}{\left(x_{2} \right)} {\operatorname{Ai}^{'}}^{2}{\left(x_{2} \right)}
\\ & \quad
+ x_{1}^{2} x_{2}^{2} \operatorname{Ai}^{4}{\left(x_{2} \right)}
+ 4 x_{1}^{2} x_{2} \operatorname{Ai}^{2}{\left(x_{2} \right)} {\operatorname{Ai}^{'}}^{2}{\left(x_{2} \right)}
+ x_{1}^{2} {\operatorname{Ai}^{'}}^{4}{\left(x_{2} \right)}
\\ & \quad
- 2 x_{1} x_{2}^{2} \operatorname{Ai}^{2}{\left(x_{2} \right)} {\operatorname{Ai}^{'}}^{2}{\left(x_{2} \right)}
- 2 x_{1} x_{2} {\operatorname{Ai}^{'}}^{4}{\left(x_{2} \right)}
+ x_{2}^{2} {\operatorname{Ai}^{'}}^4{\left(x_{2} \right)}\end{aligned}
\end{equation}

\eex

\subsection{ODE for \texorpdfstring{$W_2$}{W2} with polynomial coefficients}

In fact, we can also find an $x_1$--ODE for $W_2(X_1,X_2)$, whose coefficients are polynomial functions of both $x_1$ and $x_2$. \\
We begin by writing $\Tr M(X_1) M(X_2)$ as a quadratic form of the coefficients of $M(X_1)$ and $M(X_2)$.
Let us warm up with the example $\Lieg=\Lieg l(r,\CC)$, where $M(X_1)$ and $M(X_2)$ are $r\times r$ matrices:
\beq
W_2(X_1,X_2) = \sum_{i,j,i',j'=1}^r  M_{i',j'}(X_2) Q_{0;i',j',i,j}(x_1,x_2) M_{i,j}(X_1),
\eeq
with $Q_0(x_1,x_2)$ the $r^2\times r^2$ matrix
\beq
Q_{0;i',j',i,j}(x_1,x_2) = \frac{\delta_{i,j'}\delta_{i',j}}{(x_1-x_2)^2}.
\eeq
Hence, we get
\beq
\frac{d^k}{dx_1^k} W_2(x_1,x_2) = \sum_{i,j,i',j'=1}^r  M_{i',j'}(X_2) Q_{k;i',j',i,j}(x_1,x_2) M_{i,j}(X_1),
\eeq
with the recursion relation
\begin{equation} \label{eq:recur.rel.Qk}
\begin{aligned}
Q_{k+1;i',j',i,j}(x_1,x_2)
= \
& \frac{d}{dx_1} Q_{k;i',j',i,j}(x_1,x_2) \\
& + \sum_{l}
\left(Q_{k;i',j',l,j}(x_1,x_2) \mathcal{D}(x_1)_{l,i} - Q_{k;i',j',i,l}(x_1,x_2) \mathcal{D}(x_1)_{j,l}\right).
\end{aligned}
\end{equation}
For each $k$, $Q_k$ is an $r^2\times r^2$ matrix, and there are at most $r^4$ such matrices that can be linearly independent.
Therefore, there exist some polynomial coefficients $\tilde\alpha_{k}(x_1,x_2) $, such that
\beq
\sum_{k=0}^{r^4} \td\alpha_{k}(x_1,x_2) Q_k(x_1,x_2)=0.
\eeq
This implies
\beq
\sum_{k=0}^{r^4} \td\alpha_{k}(x_1,x_2) \frac{d^k}{dx_1^k} W_2(X_1,X_2)=0.
\eeq

\smallskip
\textbf{General case:}
Let $Q_k\in \Lieg^*\otimes \Lieg^*$ be the rational quadratic form such that
\beq
\frac{d^k}{dx_1^k} W_2(X_1,X_2) = Q_k(M(X_1),M(X_2)).
\eeq
We have the recursion
\beq
Q_{k+1}  = \frac{d}{dx_1}Q_k + [Q_k,\D(x_1)]_{1},
\eeq
where the commutator acts only on the 1st factor of $\Lieg^*\otimes \Lieg^*$ and in terms of components is given in \autoref{eq:recur.rel.Qk}.
The space $\Lieg^*\otimes \Lieg^*$ has dimension $(\dim \Lieg)^2$ and thus
there must exist some polynomial coefficients $\tilde\alpha_{k}(x_1,x_2)\in \CC[x_1,x_2] $, such that
\beq
\sum_{k=0}^{(\dim\Lieg)^2} \td\alpha_{k}(x_1,x_2) Q_k(x_1,x_2)=0.
\eeq
This implies,
\bt
$W_2$ satisfies a polynomial ODE of order $(\dim\Lieg)^2$.
There exist some coefficients $\tilde\alpha_{k}(x_1,x_2)\in \mathbb{C}[x_1,x_2]$, such that
\beq
\sum_{k=0}^{(\dim\Lieg)^2} \td\alpha_{k}(x_1,x_2) \frac{d^k}{dx_1^k} W_2(X_1,X_2)=0.
\eeq

\et

\subsection{ODEs for higher correlators}

Let $X_i=x_i.E_i$ $n$ be points in $\CC P^1\times \Lieg$.

We write
\beq
W_n(X_1,\dots,X_n) = \Tr D_{0,n}(x_1;X_2,\dots,X_n) M(X_1),
\eeq
where
\beq
D_{0,n}(x_1;X_2,\dots,X_n)
= \sum_{\sigma\in \mathfrak S_n^{1-\text{cycle}}}
(-1)^\sigma \frac{\Tr \prod_{i=1}^{n-1} M(X_{\sigma^i(1)})}{\prod_{i=1}^n (x_{\sigma(i)}-x_i)}.
\eeq
We then have
\beq
\frac{d^k}{dx_1^k} W_n(X_1,\dots,X_n) = \Tr D_{k,n}(x_1;X_2,\dots,X_n) M(X_1),
\eeq
where $D_{k,n}(x_1;X_2,\dots,X_n)$ is obtained by the recursion
\beq\label{eqdef:Dkn}
D_{k+1,n}(x_1;X_2,\dots,X_n) = \frac{d}{dx_1} D_{k,n}(x_1;X_2,\dots,X_n)
+ [D_{k,n}(x_1;X_2,\dots,X_n),\D(x_1)].
\eeq
Each $D_{k,n}(x_1;X_2,\dots,X_n)$ is rational in $x_1$ but transcendental in $x_2,\dots,x_n$.
At most $\dim\Lieg$ of them can be linearly independent, hence there exist some coefficients $\alpha_{k,n}(x_1;X_2,\dots,X_n)$ polynomial in $x_1$, such that
\beq\label{eqdef:alphakn}
\sum_{k=0}^{\dim\Lieg} \alpha_{k,n}(x_1;X_2,\dots,X_n) D_{k,n}(x_1;X_2,\dots,X_n) = 0.
\eeq
This implies
\bt
\label{thm:ODEforW_ncoef}
There exist some coefficients $\alpha_{k,n}(x_1;X_2,\dots,X_n)$, polynomial in $x_1$, such that
\beq
\sum_{k=0}^{\dim\Lieg} \alpha_{k,n}(x_1;X_2,\dots,X_n) \frac{d^k}{dx_1^k} W_n(X_1,\cdots,X_n)=0.
\eeq
The coefficients $\alpha_{k,n}$ are found by \autoref{eqdef:Dkn} and \autoref{eqdef:alphakn}.

\et

\subsection{ODE for higher correlators with polynomial coefficients} \label{sec:ODE.higher.correlators}

Moreover, there is a tensor $Q_{0,n}\in (\Lieg^*)^{\otimes n}$, such that
\beq
W_n(X_1,\dots,X_n) = Q_{0,n}(M(X_1),M(X_2),\dots,M(X_n))
\eeq
whose coefficients are rational functions of $x_1,\dots,x_n$. For $\Lieg l(r,\CC)$, 
the expression of $Q_{0,n}$ can be written as:
\beq \label{eq:Q0n}
Q_{0,n}=\sum_{\sigma\in \mathfrak S_n^{1-cycle}} (-1)^{\sigma}\sum_{l_1,\ldots ,l_n} \frac{e^{1}_{l_1l_{\sigma(1)}}\otimes \cdots \otimes e^{n}_{l_{n}l_{\sigma(n)}}}{\prod_i(x_{\sigma(i)}-x_i)},
\eeq
with $e_{i,j} \in \Lieg l(r,\CC)^*$ and satisfy $e_{i,j}(M)=M_{i,j}$ for any $M\in \Lieg l(r,\CC)$.
Moreover, we get
\beq
\frac{d^k}{dx_1^k} W_n(X_1,\dots,X_n) = Q_{k,n}(M(X_1),M(X_2),\dots,M(X_n)),
\eeq
where
\beq\label{eqdef:Qknrec}
Q_{k+1,n} = \frac{d}{dx_1} Q_{k,n} + [Q_{k,n},\D(x_1)]_{1}.
\eeq
The commutator
acts only on the 1st factor of $({\Lieg^*})^{\otimes n}$.
The space $({\Lieg^*})^{\otimes n}$ has dimension $(\dim\Lieg)^n $, and so at most $(\dim\Lieg)^n$ such tensors can be linearly independent.
Consequently, there exist some polynomial coefficients
$\td\alpha_{k,n}(x_1,\dots,x_n)\in \CC[x_1,\dots,x_n]$, such that
\beq\label{eqdef:tdalphakn}
\sum_{k=0}^{\dim\Lieg^n} \td\alpha_{k,n}(x_1,\dots,x_n) \ Q_{k,n}(x_1;x_2,\dots,x_n)=0.
\eeq

This implies
\bt
\label{thm:ODEforW_ncoefs}
There exist some coefficients $\tilde\alpha_{k,n}(x_1;x_2,\dots,x_n) \in \mathbb{C}[x_1,\ldots,x_n]$, such that
\beq
\sum_{k=0}^{\dim\Lieg^n} \td\alpha_{k,n}(x_1,\dots,x_n) \frac{d^k}{dx_1^k} W_n(X_1,X_2,\dots,X_n)=0.
\eeq
The coefficients $\td\alpha_{k,n}$ are found by \autoref{eqdef:Qknrec} and \autoref{eqdef:tdalphakn}.
The $(\dim\Lieg)^n$ linearly independent solutions are obtained by choosing $(E_1,E_2,\dots,E_n)\in\Lieg^{\otimes n}$.
\et

\section{Differential equations with respect to other parameters}
\label{sec:genPDE}

Consider a compatible \textbf{family} of such systems, parametrized by times $\mathbf t=(t_1,t_2,\dots)$, i.e. we have a time-dependent family (locally $C^\infty$) of sections $\Psi(\mathbf t;x)$ of a Lie group $G$ principal bundle over $\CC P^1$ with rational connections:
\beq
\frac{d}{dx} \Psi(\mathbf t;x) = \D(\mathbf t;x) \Psi(\mathbf t;x),
\eeq
\beq
\frac{d}{dt_k} \Psi(\mathbf t;x) = R_k(\mathbf t;x) \Psi(\mathbf t;x),
\eeq
where each $R_k(\mathbf t;x)$ is rational in $x$ (but can be a transcendantal function of $\mathbf t$).

The compatibility of these equations is equivalent to saying that there exists (locally) a common solution (locally analytic), which itself is equivalent to the Lax equations:
\beq
\frac{d}{dt_k} \D(\mathbf t;x) = [R_k(\mathbf t;x),\D(\mathbf t;x)] + \frac{d}{dx}R_k(\mathbf t;x),
\eeq
\beq
\frac{d}{dt_k} R_l(\mathbf t;x) = [R_k(\mathbf t;x),R_l(\mathbf t;x)] + \frac{d}{dt_l}R_k(\mathbf t;x).
\eeq

The adjoint flat sections are defined as in  \autoref{def:AdjointFlatSection} by replacing $\Psi(x) \rightarrow \Psi(\mathbf t;x)$
\beq
M(\mathbf t;x.E) = \Psi(\mathbf t;x) E \Psi(\mathbf t;x)^{-1},
\eeq
with $E\in \Lieg$, independent of $x$ and $\mathbf t$.

They satisfy the adjoint ODE
\beq
\frac{d}{dt_j} M(\mathbf t;x.E) = [R_j(\mathbf t;x),M(\mathbf t;x.E)].
\eeq

\subsection{Time Differential equation  for \texorpdfstring{$W_1$}{W1}}\label{sec:time.diff.eqt.W1}

Let us choose a time $t_j$. For each $k\geq 0$ we can write
\beq
\frac{d^k}{dt_j^k} W_1(x.E) = \Tr R_{j,k}(\mathbf t;x) \ M(\mathbf t;x.E),
\eeq
where the matrix $R_{j,k}(\mathbf t;x)$ is defined by the following recursion:
\beq \label{eqdef:Rjk}
\begin{aligned}
\text{initial term:}&\qquad && R_{j,0}(\mathbf t;x) = \D(\mathbf t;x)
\\
\text{recursion:}&\quad&&
R_{j,k+1}(\mathbf t;x) = \frac{d}{dt_j} R_{j,k}(\mathbf t;x) + [R_{j,k}(\mathbf t;x),R_j(\mathbf t;x)]
\end{aligned}
\eeq
Again at most $\dim\Lieg$ of them are independent, there must exist some coefficients polynomial in $x$ such that
\beq\label{eqdef:alphajk1}
\sum_{k=0}^{\dim\Lieg} \alpha_{j,k,1}(\mathbf t;x) R_{j,k}(\mathbf t;x) =0.
\eeq
This implies
\bt[Time differential equation for $W_1$]
there exist coefficients $\alpha_{j,k,1}(\mathbf t;x)$ polynomial in $x$, such that
\beq
\sum_{k=0}^{\dim\Lieg} \alpha_{j,k,1}(\mathbf t;x) \frac{d^k}{dt_j^k}W_1(\mathbf t;x.E) =0.
\eeq
The coefficients $\alpha_{j,k,1}$ are determined by \autoref{eqdef:Rjk} and \autoref{eqdef:alphajk1}.
\et

\subsection{Time Differential equation for \texorpdfstring{$W_n\,, n \geq 2$}{higher correlators}}

Results of \autoref{sec:time.diff.eqt.W1} directly generalise to the case of n-point correlators for $n \geq 2$

\bt
there exist coefficients $\alpha_{j,k,n}(\mathbf t;x_1;X_2,\dots,X_n)$ polynomial in $x_1$, such that
\beq
\sum_{k=0}^{\dim\Lieg} \alpha_{j,k,n}(\mathbf t;x_1;X_2,\dots,X_n) \frac{d^k}{dt_j^k}W_n(\mathbf t;X_1,X_2,\dots,X_n) =0,
\eeq
where $\alpha_{j,k,n}$ are determined by
\beq\label{eqdef:Rjkn}
\begin{aligned}
\text{Initial term:}& \qquad && R_{j,0,n}(\mathbf t;x_1,X_2,\dots,X_n) = D_{0,n}(\mathbf t;x_1,X_2,\dots,X_n). \\
\text{Recursion:}& \quad &&
R_{j,k+1,n}(\mathbf t;x_1,X_2,\dots,X_n) = \frac{d}{dt_j} R_{j,k,n}(\mathbf t;x_1,X_2,\dots,X_n) \\
& && \qquad \qquad + [R_{j,k,n}(\mathbf t;x_1,X_2,\dots,X_n),R_j(\mathbf t;x_1)]. \\
\text{Linear relation:}& \quad && \sum_{k=0}^{\dim\Lieg} \alpha_{j,k,n}(\mathbf t;x_1,X_2,\dots,X_n) R_{j,k,n}(\mathbf t;x_1,X_2,\dots,X_n) =0,
\end{aligned}
\eeq
where
\beq
D_{0,n}(x_1;X_2,\dots,X_n)
= \sum_{\sigma\in \mathfrak S_n^{1-\text{cycle}}}
(-1)^\sigma \frac{\Tr \prod_{i=1}^{n-1} M(X_{\sigma^i(1)})}{\prod_{i=1}^n (x_{\sigma(i)}-x_i)}.
\eeq
\et

\bt
there exist coefficients $\tilde\alpha_{j,k,n}(\mathbf t;x_1;x_2,\dots,x_n) \in \mathbb{C}[x_1, \ldots, x_n]$, such that
\beq
\sum_{k=0}^{\dim\Lieg} \td\alpha_{j,k,n}(\mathbf t;x_1;x_2,\dots,x_n) \frac{d^k}{dt_j^k}W_n(\mathbf t;x_1.E_1,x_2.E_2,\dots,x_n.E_n) =0.
\eeq
where $\td\alpha_{j,k,n}$ are determined by
\beq
\begin{aligned}
\text{Initial term:}& \qquad && \td Q_{j,0,n}(\mathbf t;x_1,\dots,x_n) = Q_{0,n}(\mathbf t;x_1,\dots,x_n). \\
\text{Recursion:}& \quad &&
\td Q_{j,k+1,n}(\mathbf t;x_1,\dots,x_n) = \frac{d}{dt_j} \td Q_{j,k,n}(\mathbf t;x_1,\dots,x_n) \\
& &&\qquad \qquad + \sum_{i=1}^n [\td Q_{j,k,n}(\mathbf t;x_1,\dots,x_n),R_j(\mathbf t;x_i)]_{i}. \\
\text{Linear relation:}& \quad && \sum_{k=0}^{\dim\Lieg} \td\alpha_{j,k,n}(\mathbf t;x_1,x_2,\dots,x_n) \td Q_{j,k,n}(\mathbf t;x_1,\dots,x_n) =0,
\end{aligned}
\eeq
where
\beq
Q_{0,n}=\sum_{\sigma\in \mathfrak S_n^{1-cycle}} (-1)^{\sigma}\sum_{l_1,\ldots ,l_n} \frac{e^{1}_{l_1l_{\sigma(1)}}\otimes \cdots \otimes e^{n}_{l_{n}l_{\sigma(n)}}}{\prod_i(x_{\sigma(i)}-x_i)}.
\eeq
\et

\section{Recursion relations}
\label{sec:genRec}

Now consider a compatible \textbf{sequence}   of such systems indexed by an integer $N$, i.e. we have a section $\Psi^{(N)}(x)$ of a Lie group $G$ principal bundle over $\CC P^1$ with rational connections:
\beq
\label{eq:comp1}
\frac{d}{dx} \Psi^{(N)}(x) = \D^{(N)}(x) \Psi^{(N)}(x),
\eeq
that satisfies a recursion
\beq
\label{eq:comp2}
\Psi^{(N+1)}(x) = \mathcal R^{(N)}(x) \Psi^{(N)}(x),
\eeq
where each $ \mathcal R^{(N)}(x)$ is rational in $x$ (but can be any function of $N$).

In analogy to \autoref{sec:genPDE}, the existence of a common solution (locally) to eq's \eqref{eq:comp1}-\eqref{eq:comp2} implies a compatibility relation called the discret Lax equations, namely:

\bp[Compatibility]
The ODE and recursion are compatible:
\beq
\mathcal D^{(N+1)}(x)\mathcal R^{(N)}(x) - \mathcal R^{(N)}(x) \mathcal D^{(N)}(x)
= \frac{d}{dx} \mathcal R^{(N)}(x).
\eeq
\ep
We write it as
\beq
\begin{aligned}
\mathcal D^{(N+1)}(x)
&=&& \mathcal R^{(N)}(x) \mathcal D^{(N)}(x)\mathcal R^{(N)}(x)^{-1}
+ \frac{d}{dx}\mathcal R^{(N)}(x) \ \mathcal R^{(N)}(x)^{-1} \cr
&=&& \mathcal R^{(N)}(x) \left( \mathcal D^{(N)}(x)
+ \mathcal R^{(N)}(x)^{-1}\frac{d}{dx}\mathcal R^{(N)}(x) \right) \mathcal R^{(N)}(x)^{-1} \cr
&=&& \mathcal R^{(N)}(x) \  \mathcal D^{(N,1)}(x)
\ \mathcal R^{(N)}(x)^{-1},
\end{aligned}
\eeq
where we defined
\beq
\mathcal D^{(N,1)}(x) = \mathcal D^{(N)}(x)
+ \mathcal R^{(N)}(x)^{-1}\frac{d}{dx}\mathcal R^{(N)}(x) .
\eeq

We also define the adjoint flat sections
\beq
M^{(N)}(x.E) = \Psi^{(N)}(x) E \Psi^{(N)}(x)^{-1}. 
\eeq
\bp[Recursion and ODE for the adjoint section]\label{prop:Rec.and.ODE.adjoint}
We have:
\beq
M^{(N+1)}(x) = \mathcal R^{(N)}(x) M^{(N)}(x) \mathcal R^{(N)}(x)^{-1},\quad
\frac{d}{dx} M^{(N)}(x) = [\mathcal D^{(N)}(x), M^{(N)}(x)].
\eeq

\ep


\subsection{Recursion for \texorpdfstring{$W^{(N)}_1$}{W1} }

We have
\beq
W_1^{(N)}(x.E) = \Tr \D^{(N)}(x) M^{(N)}(x.E),
\eeq
and thus
\beq
\begin{aligned}
W_1^{(N+1)}(x.E)
&=&& \Tr \D^{(N+1)}(x) M^{(N+1)}(x.E) \cr
&=&& \Tr \D^{(N+1)}(x) \mathcal R^{(N)}(x) M^{(N)}(x.E) \mathcal R^{(N)}(x)^{-1} \cr
&=&& \Tr \mathcal R^{(N)}(x)^{-1} \D^{(N+1)}(x) \mathcal R^{(N)}(x) M^{(N)}(x.E) \cr
&=&& \Tr \left(  \D^{(N)}(x) +\mathcal R^{(N)}(x)^{-1} \mathcal R'^{(N)}(x) \right) M^{(N)}(x.E).
\end{aligned}
\eeq
Notice that,
$
W_1^{(N+1)}(x.E) - W_1^{(N)}(x.E)
=
\Tr \mathcal R^{(N)}(x)^{-1} \mathcal R'^{(N)}(x) \  M^{(N)}(x.E)
$.

Then, in general we shall have
\bl[$W_1^{(N+k)}(x.E)$]
\beq
W_1^{(N+k)}(x.E) =
\Tr D_k^{(N)}(x) \ M^{(N)}(x.E),
\eeq
with $D_0^{(N)}(x) = \D^{(N)}(x)$ and,
\beq
D_k^{(N)}(x) = \mathcal R^{(N)}(x)^{-1} \cdots \mathcal R^{(N+k-1)}(x)^{-1} \ \ \D^{(N+k)}(x)  \ \ \mathcal R^{(N+k-1)}(x) \cdots  \mathcal R^{(N)}(x)
\eeq
is rational in $x$.
\el

At most $\dim\Lieg$ of these matrices can be independent, and therefore, there exist 	 some coefficients $\alpha_k^{(N)}(x) \in \mathbb{C}[x]$, such that:
\beq
\sum_{k=0}^{\dim\Lieg} \alpha_k^{(N)}(x) D_k^{(N)}(x)=0.
\eeq
This implies
\bt[Recursion equation for $W_1$]\label{thm:Rec.W1}
$W_1(x.E)$ satisfies a  linear recursion relation of order $\dim\Lieg$, with polynomial coefficients $\alpha_k^{(N)}(x) $, independent of $E$:
\beq
\sum_{k=0}^{\dim\Lieg} \alpha_k^{(N)}(x) W_1^{(N+k)}(x.E)=0.
\eeq

\et

\subsection{Recursion for \texorpdfstring{$W^{(N)}_n, \ n \geq 2$}{higher correlators}}

The rest proceeds in the same way as for differential equations:

\bt[Recursion equation for $W^{(N)}_n$]\label{thm:Rec.for.Wn}
$W^{(N)}_n(x_1.E_1,\dots,x_n.E_n)$ satisfies a  linear recursion relation of order $(\dim\Lieg)^n$, with coefficients $\td\alpha_k^{(N)}(x_1,\dots,x_n) \in \mathbb{C}[x_1,\ldots,x_n]$, independent of $E_1,\dots, E_n$:
\beq
\sum_{k=0}^{(\dim\Lieg)^n} \td\alpha_{k,n}^{(N)}(x_1,\dots,x_n) W_n^{(N+k)}(x_1.E_1,\dots,x_n.E_n)=0,
\eeq

where $\td\alpha_{k,n}^{(N)}$ are determined by:
\beq
\begin{aligned}
\text{Initial term}:& \qquad &&\td Q^{(N)}_{0,n}(x_1,\dots,x_n)  = Q_{0,n}(x_1,\dots,x_n).
\\
\text{Recursion}:& \qquad && \td Q^{(N)}_{k+1,n}(x_1,\dots,x_n)
=
\\
& &&
\left(
\left(
\Ad_{ \mathcal R^{(N+k)}(x_1)}\otimes \dots \otimes \Ad_{ \mathcal R^{(N+k)}(x_n)}  
\right)
\td Q^{(N)}_{k,n}
\right)
(x_1,\dots,x_n) .
\\
\text{Linear relation}:& \qquad && \sum_{k=0}^{(\dim\Lieg)^n} \td\alpha_{k,n}^{(N)}(x_1,\dots,x_n)  \td Q^{(N)}_{k,n}(x_1,\dots,x_n)=0.
\end{aligned}
\eeq
\et
Notice that the initial term $\td Q^{(N)}_{0,n}(x_1,\dots,x_n) = Q_{0,n}(x_1,\dots,x_n)$ is actually independent of $N$.

\proof
By definition $W_n^{(N+k)}(x_1,\dots,x_n) = \td Q_{k,n}^{(N)}(M(x_1.E_1),\dots,M(x_n.E_n))$. Therefore, proving the recursion and linear relation will conclude the proof. 

The tensor $Q_{0,n}$ is given by the application: 
\begin{align}
Q_{0,n} \colon \Lieg^{\otimes n} &\to \CC \\
(M_1,...,M_n) &\mapsto \sum_{\sigma\in \mathfrak S_n^{1-\text{cycle}}}
(-1)^\sigma \frac{\Tr \prod_{i=1}^n M_{\sigma^i(1)}}{\prod_{i=1}^n (x_{\sigma(i)}-x_i)}
\end{align}
Due to \autoref{def:Wngeneral} and \autoref{prop:Rec.and.ODE.adjoint} we get,
\begin{align}
W_n^{(N+1)}\ &=Q_{0,n}(M^{(N+1)}(X_1),\cdots,M^{(N+1)}(X_n))
\\
&=Q_{0,n}(\Ad_{\mathcal R^{(N)}(x_1)}M^{(N)}(X_1),\cdots,\Ad_{\mathcal R^{(N)}(x_n)}M^{(N)}(X_n))
\\
&\eqqcolon 
\left(
\left(
\Ad_{\mathcal R^{(N)}(x_1)}\otimes \dots \otimes \Ad_{ \mathcal R^{(N)}(x_n)}
\right)
Q_{0,n}
\right)
(M^{(N)}(X_1),\cdots,M^{(N)}(X_n)). 
\end{align}
By induction, we conclude that,
\beq
\td Q^{(N)}_{k+1,n}(x_1,\dots,x_n) 
=
\left(
\left(
\Ad_{ \mathcal R^{(N+k)}(x_1)}\otimes \dots \otimes \Ad_{ \mathcal R^{(N+k)}(x_n)} \right)
\td Q^{(N)}_{k,n}\right)(x_1,\dots,x_n).
\eeq
Finally, since $ Q^{(N)}_{k,n} \in (\Lieg^*)^{\otimes n}$ at most $(\dim \Lieg)^n$ of them can be linearly independent. Therefore, there exist coefficients $\td\alpha_{k,n}^{(N)} \in \mathbb{C}[x_1,\ldots,x_n]$ such that
\beq
\sum_{k=0}^{(\dim\Lieg)^n} \td\alpha_{k,n}^{(N)}(x_1,\dots,x_n)  \td Q^{(N)}_{k,n}(x_1,\dots,x_n)=0.
\eeq

\eproof		

Remark, that in the case of $\Lieg = \Lieg l(r,\CC)$ the presented proof for the recursion could be formulated using index notation, \autoref{eq:Q0n} reads
\begin{equation}
Q_{0,n}=\sum_{\sigma\in \mathfrak S_n^{1-cycle}} (-1)^{\sigma}\sum_{l_1,\ldots,l_n} \frac{e^{1}_{l_1l_{\sigma(1)}}\otimes \cdots \otimes e^{n}_{l_{n}l_{\sigma(n)}}}{\prod_i(x_{\sigma(i)}-x_i)}.
\end{equation}

 Moreover, with the use of \autoref{def:Wngeneral} and \autoref{prop:Rec.and.ODE.adjoint}
\begin{equation}
\begin{aligned}
W_n^{(N+1)} \
&=
\sum_{\sigma\in \mathfrak S_n^{1-\text{cycle}}}
(-1)^\sigma \frac{\Tr \prod_{i=1}^n M^{(N+1)}(x_{\sigma^i(1)}.E_{\sigma^i(1)})}{\prod_{i=1}^n (x_{\sigma(i)}-x_i)}
\\
&=
\sum_{\sigma\in \mathfrak S_n^{1-cycle}}(-1)^{\sigma} \frac{\Tr \left(\prod_i  \mathcal R^{(N)}(x_{\sigma^i(1)})M^{(N)}(x_{\sigma^i(1)}) \mathcal R^{(N)}(x_{\sigma^i(1)})^{-1}\right)}{\prod_i(x_{\sigma(i)}-x_i)}
\\
&=
\left(
Q_{0,n}\rvert_{e^{i}_{l_{i},l_{\sigma(i)}} \to \Ad_{ \mathcal R^{(N)}\left(x_{i}\right)} e^{i}_{l_{i},l_{\sigma(i)}}}
\right)
(M^{(N)}(x_1),\ldots,M^{(N)}(x_N))
\end{aligned}
\end{equation}

which leads to 
\begin{equation}
\tilde Q^{(N)}_{1,n}
= 
\left(
\left(
\Ad_{\mathcal R^{(N)}(x_1)}\otimes \dots \otimes \Ad_{ \mathcal R^{(N)}(x_n)}  
\right)
Q_{0,n}\right)(x_1,\dots,x_n).
\end{equation}
Finally, $\tilde Q^{(N)}_{k,n}$ is obtained by induction. 

\section{Applications}
\subsection{1-Matrix model}
\label{sec:1MM}

Expectation values of spectral densities of random matrices, typically fit in the framwork developed here.
So let us see how it applies to random matrices to get differential equations and recursions.

\subsubsection{Introduction}

Usually, random matrix theory is defined with a probability law on a set of Hermitian (resp. Normal) matrices with a measure continuous with respect to the Lebesgue measure.

However, since most of the time the random variables of interest concern the eigenvalues, we may also directly define random matrix theory as its marginal law of eigenvalues.

The eigenvalues induced measure (possibly complex measure) takes the following form for the diagonal matrix $\Lambda$ of eigenvalues:
\beq
d\mu(\Lambda) =  \frac{1}{N! Z} \ \Delta(\Lambda)^2 e^{-\Tr V(\Lambda)} d\Lambda,\quad \Lambda = \diag(\lambda_1,\lambda_2,\dots,\lambda_N) \in \gamma^N,
\eeq
where the function $V$ is called the potential, $d\Lambda$ is the product measure $d\Lambda=d\lambda_1\dots d\lambda_N$  on $\gamma^N$,
and the Vandermonde $\Delta(\Lambda)$ is defined as
\beq
\Delta(\Lambda) = \prod_{i<j} (\lambda_i-\lambda_j),
\eeq
and the partition function $Z$ is
\beq
Z =  \frac{1}{N!} \int_{\gamma^N} \ \Delta(\Lambda)^2 e^{-\Tr V(\Lambda)} d\lambda_1\dots d\lambda_N.
\eeq
The integration domain $\gamma$ is the space to which the $\lambda_i$s belong.
It needs not be the real axis, it may be any space on which this measure can be defined, it could be for example a Polish subspace of $\RR$, but it can also be a Jordan arc in the complex plane, as long as $Z$ is an absolutely convergent integral.

\bex[Gaussian Unitary Ensemble (GUE)]
The Gaussian Unitary Ensemble is the case $\gamma=\RR$ and $V(x)=\frac12 x^2$. It is, thus, a probability law on $\RR^N$:
\beq
d\mu(\Lambda) = \frac{1}{Z^{(N)}} \ \Delta(\Lambda)^2 \ e^{-\frac12 \Tr \Lambda^2} d\Lambda,
\eeq
$Z^{(N)}$ is known to be proportional to the Barnes function
\beq
Z^{(N)} = (2\pi)^{\frac N 2} \ \prod_{k=0}^{N-1} k!.
\eeq

\eex

\subsubsection{Potentials and Jordan arcs}

It is customary to consider potentials to be polynomials or rational, and to choose $\gamma$ a Jordan arc such that the measure vanishes at the boundaries of $\gamma$ (the boundaries may be at $\infty$).

\textbf{Polynomial case}:

Let $V(x)\in \CC[x]$ be a polynomial with $\deg V\geq 1$ called the potential.
Notice that $e^{-V(x)}$ can vanish only at $\infty$.

An integral
\beq
\int_\gamma f(x) e^{-V(x)}dx,
\eeq
where $f(x)\in \CC[x]$ is invariant under homotopic deformations of the smooth Jordan arc $\gamma$, and it can be absolutely convergent only if $\gamma$ goes to $\infty$ in directions in which $\Re V(x)\to +\infty$.

There are exactly $\deg V$ angular sectors around $\infty$ in which $\Re V(x)\to +\infty$, and thus there are $\deg V -1=\deg V'$ independent possible Jordan arcs (up to addition by concatenation), that go from an allowed sector to another one.
We shall say that the homology space of ``admissible" paths is of dimension
\beq
\dim (\text{Homology space of admissible paths}) = \deg V'.
\eeq

If $V(x)$ is an even degree real polynomial whose leading coefficient is $>0$, then in that case, $\gamma=\RR$ is an admissible path.

For example if $V(x) = \frac12 x^2$, then $\deg V'=1$ and thus $\gamma=\RR$ is the only (up to homotopy and multiplication by a non-zero scalar) admissible path.

Otherwise admissible paths are often complex paths.

\textbf{Rational case}:

We repeat the same considerations if $V'(x)\in \CC(x)$ is a rational function. This means that $V(x)$ may have logarithmic singularities. $e^{-V}$ may vanish only at the poles of $V'$.

An integral
\beq
\int_\gamma f(x) e^{-V(x)}dx,
\eeq
where $f(x)\in \CC[x]$ is invariant under homotopic deformations of the smooth Jordan arc $\gamma$, and it can be convergent only if $\gamma$ approaches singularities of $V$ in directions in which $\Re V(x)\to +\infty$.

It is explained in \cite{BertolaEynard} how to construct the homology space of all admissible paths, and its dimension is found to be
\beq
\dim = \deg V' = \text{sum of degrees of all poles, including the pole at }\infty.
\eeq

%

\subsubsection{Definitions}

\textbf{Wave functions  and orthogonal polynomials}:

\bd[Wave functions]
Let
\beq
p^{(N)}(x) = \int_{\gamma^N} d\mu(\Lambda) \ \det(x-\Lambda).
\eeq
It is a polynomial of $x$ of degree $N$ (it is the expectation value of the characteristic polynomial).
\ed

\bp[Orthogonal polynomials]
The polynomials $p^{(N)}$ form a family of orthogonal polynomials for the measure $e^{-V(x)}dx$ on $\gamma$:
\beq
\int_\gamma  dx \ p^{(N)}(x) \  p^{(N')}(x)e ^{-V(x)} = h^{(N)} \ \delta_{N,N'}.
\eeq
\ep
\proof
It is a well known result \cite{Heine, Szego, MehtaBook}
\eproof

\bd[Normalized wave functions]
We thus define
\beq
\psi^{(N)}(x) = \frac{e^{-\frac12 V(x)}}{\sqrt{h^{(N)}}} \ p^{(N)}(x),\quad h^{(N)} = \int_\gamma dx \ p^{(N)}(x) p^{(N)}(x) e^{-V(x)}, \quad
\gamma_N = \sqrt{\frac{h^{(N)}
}{h^{(N-1)}}}.
\eeq
They form an orthonormal $L_2(\gamma)$ family:
\beq
\int_\gamma dx \ \psi^{(N)}(x) \psi^{(N')}(x)  = \delta_{N,N'}.
\eeq
\ed

\bd[Dual wave functions]
We also define
\beq
\phi^{(N)}(x) =  \frac{e^{\frac12 V(x)}}{\sqrt{h^{(N)}}} \int_\gamma  dx' e^{-V(x')} \frac{p^{(N)}(x')}{x-x'}.
\eeq
\ed
\bd[Matrix of wave functions] The matrix wave function is defined as:
\beq
\Psi^{(N)}(x) = \begin{pmatrix}
\psi^{(N-1)}(x) & \phi^{(N-1)}(x)  \cr
\psi^{(N)}(x) & \phi^{(N)}(x)  \cr
\end{pmatrix}.
\eeq

\ed

\bex[Gaussian  (GUE)] For the Hermite polynomial, $p^{(N)}(x) = H_N(x)$, we have:
\beq
\int_\RR H_N(x) H_M(x) e^{-\frac12 x^2} dx = \sqrt{2\pi} N! \ \delta_{N,M}, \quad
h^{(N)} = \sqrt{2\pi} N!, \quad
\gamma^{(N)}=\sqrt{N}.
\eeq

\eex

\subsubsection{Resolvents}

\bd[Resolvents]
Let
\beq
W_1^{(N)}(x)
= \int_{\gamma^N} d\mu(\Lambda) \  \Tr  \frac{1}{x-\Lambda},
\eeq
and in general
\beq
\hat{W}_n^{(N)}(x_1,\dots,x_n) = \frac{ \delta_{n,2}}{(x_1-x_2)^2}+  \int_{\gamma^N}  \ d\mu(\Lambda) \prod_{i=1}^n \Tr \frac{1}{x_i-\Lambda}.
\eeq
Let $W_n$ be their cumulants, defined by inverting the formula:
\beq\label{eqdef:cumulantsWn}
\hat{W}_n^{(N)}(x_1,\dots,x_n)
= \sum_{\mu \vdash \{x_1,\dots,x_n\}} \prod_{i=1}^{\ell(\mu)} W^{(N)}_{|\mu_i|}(\mu_i),
\eeq
i.e. the $\hat{W}_n^{(N)}$ are the sums over all partitions of the $n$ variables, of products of cumulants of each part.

\ed
The system of equations \eqref{eqdef:cumulantsWn} that define the cumulants is a triangular system easy to invert, it gives for example $W_1^{(N)}=\hat W_1^{(N)}$, and the $n=2$ cumulant is the covariance of resolvents:
\beq
W_2^{(N)}(x_1,x_2)= \hat W_2^{(N)}(x_1,x_2) - W_1^{(N)}(x_1) W_1^{(N)}(x_2).
\eeq
Similarly
\beq
\begin{aligned}
W_3^{(N)}(x_1,x_2,x_3)
&=&& \hat W_3^{(N)}(x_1,x_2,x_3) - W_1^{(N)}(x_1) \hat W_2^{(N)}(x_2,x_3) \cr
& && - W_1^{(N)}(x_2) \hat W_2^{(N)}(x_1,x_3) - W_1^{(N)}(x_3) \hat W_2^{(N)}(x_1,x_2) \cr
& && + 2 W_1^{(N)}(x_1) W_1^{(N)}(x_2)  W_1^{(N)}(x_3) .
\end{aligned}
\eeq

\subsubsection{Spectral densities and correlations}\label{sec:1MMdensities}

\bd[Spectral density and correlations]
Let $x\in\gamma$, define the ``expected spectral density"
\beq
\rho_1^{(N)}(x)
= \int_{\gamma^N} d\mu(\Lambda) \  \sum_{i=1}^N \delta(x-\lambda_i).
\eeq
In other words:
\begin{center}
$\frac{1}{N}\rho_1^{(N)}(x)$ is the probability that $x$ is an eigenvalue of $\Lambda$.
\end{center}
and in general for $x_1,\dots,x_n\in \gamma^n $, define
\beq
\hat{\rho}_n^{(N)}(x_1,\dots,x_n) =   \int_{\gamma^N}  \ d\mu(\Lambda) \prod_{i=1}^n \left(\sum_{j=1}^N \delta(x_i-\lambda_j) \right).
\eeq
Let $\rho_n$ be their cumulants, defined by the formula:
\beq\label{eqdef:cumulantsrhon}
\hat{\rho}_n^{(N)}(x_1,\dots,x_n)
= \sum_{\mu \vdash \{x_1,\dots,x_n\}} \prod_{i=1}^{\ell(\mu)} \rho^{(N)}_{|\mu_i|}(\mu_i),
\eeq
i.e. the $\hat{\rho}_n^{(N)}$ are the sums over all partitions of the $n$ variables of products of cumulants of each part.

In other words:
$\frac{1}{N^n}\hat{\rho}_n^{(N)}(x_1,\dots,x_n)$ is the expectation that $x_1,\dots,x_n$ are simultaneously eigenvalues, it is interpreted as the correlation function of $n$ eigenvalues.

The cumulants $\frac{1}{N^n}{\rho}_n^{(N)}(x_1,\dots,x_n)$ are called the connected correlation function of $n$ eigenvalues, they generalize the covariance for $n\geq 3$.

\ed

Resolvents and spectral correlations are closely related.

\begin{itemize}

\item Resolvents are ``Stieltjes transforms" of spectral correlations:
\beq
W_1^{(N)}(x) = \int_{x'\in \gamma} \frac{1}{x-x'} \ \rho_1^{(N)}(x')dx',
\eeq
\beq
W_n^{(N)}(x_1,\dots,x_n) = \int_{\gamma^n} \prod_{i=1}^n\frac{1}{x_i-x'_i} \ \rho_n^{(N)}(x'_1,\dots,x'_n) dx'_1\dots dx'_n.
\eeq

\item Spectral correlations are discontinuities of resolvents.
Notice that the resolvents are ill-defined whenever some $x_i$'s belong to the Jordan arc $\gamma$.
In fact they have a discontinuity across $\gamma$, and this discontinuity is proportional to the spectral density:
\beq
\rho_1^{(N)}(x) = \frac{1}{2\pi i} \left(W^{(N)}_1(x_\text{left})-W^{(N)}_1(x_\text{right}) \right) = \frac{1}{2\pi i} \operatorname{Disc} W^{(N)}_1(x),
\eeq
\beq\label{eq:spectradensitydiscWn}
\rho_n^{(N)}(x_1,\dots,x_n) = \frac{1}{(2\pi i)^n}  \operatorname{Disc}_{x_1}\dots \operatorname{Disc}_{x_n} W^{(N)}_n(x_1,\dots,x_n).
\eeq

\end{itemize}

\bc[Same linear ODE and recursions]\label{cor:sameODEWrho}
As a consequence of \autoref{eq:spectradensitydiscWn}, if $W_n$ satisfies a linear relation (ODE, PDE or recursion) with rational coefficients (in the variables $x,x_1,\dots,x_n$), then the spectral correlation $\rho_n$ satisfy the same linear  relation.

\ec

\subsubsection{Determinantal formulae}

The following relation is well known \cite{BE09}, \cite[p.~162-163]{bookEynardKimuraRibault}
\bp[Determinantal formulae]

Let
\beq
E=\begin{pmatrix}
1 & 0 \cr 0 & 0
\end{pmatrix}, \quad
M^{(N)}(x) = \Psi^{(N)}(x) E \Psi^{(N)}(x)^{-1}.
\eeq

We have
\beq
W_1^{(N)}(x) = \frac{V'(x)}{2}+\Tr \Psi^{(N)'}(x) E \Psi^{(N)}(x)^{-1},
\eeq
\beq
W_2^{(N)}(x_1,x_2) = \frac{1}{(x_1-x_2)^2} \Tr M^{(N)}(x_1) M^{(N)}(x_2)-\frac{1}{(x_1-x_2)^2},
\eeq
and for $n\geq 3$ we have
\beq
W_n^{(N)}(x_1,\dots,x_n)  = W_n(x_1.E,\dots,x_n.E),
\eeq
where the right hand side is ~\autoref{eq:defWn} of ~\autoref{def:Wngeneral}.

\ep

\br
In fact, the initial motivation of \cite{BE09} for introducing all this formalism, was that it applied to random matrices. So this proposition was actually not a consequence, but the origin.
\er

\subsubsection{Recursion and ODE for the wave functions}

\textbf{3-terms recursion relations}:

The following is well known  \cite{Szego,MehtaBook}
\bp[Jacobi 3-term recursion relation]
The orthogonal polynomials satisfy a 3 terms recursion relation
\beq
x p^{(N)}(x) = p^{(N+1)}(x) + S_N p^{(N)}(x) + \frac{h^{(N)}}{h^{(N-1)}} p^{(N-1)}(x),
\eeq
which gives
\beq
x \psi^{(N)}(x) = \gamma^{(N+1)} \psi^{(N+1)}(x) + S_N \psi^{(N)}(x) + \gamma^{(N)} \psi^{(N-1)}(x).
\eeq
We write it:
\beq
x \psi^{(N)}(x) =  \sum_{m\in \mathbb N} Q_{N,m} \psi^{(m)}(x),
\eeq
where $Q$ is an infinite matrix, with only 3 bands:
\beq\label{eq:defQ1MM}
Q_{n,m} = S_n \delta_{n,m} + \gamma^{(n)} \delta_{n,m+1} + \gamma^{(m)}\delta_{n+1,m}.
\eeq
It is symmetric, $Q_{n,m}=Q_{m,n}$ i.e.
\beq
Q= Q^T.
\eeq
In particular, $S_N=Q_{N,N}$. 
It implies a recursion for the matrix $\Psi$:
\beq
\Psi^{(N+1)}(x) = \mathcal R^{(N)}(x) \Psi^{(N)}(x),
\eeq
where
\beq
\mathcal R^{(N)}(x) = \begin{pmatrix}
0 & 1 \cr
\frac{-\gamma^{(N)}}{\gamma^{(N+1)}} & \frac{x-S_N}{\gamma^{(N+1)}}
\end{pmatrix}
\eeq
is called the shift operator, or sometimes the ``ladder" operator.

$\mathcal R^{(N)}(x)$ is a matrix polynomial of $x$ of degree 1.
Its inverse is also polynomial of degree 1:
\beq
\mathcal R^{(N)}(x)^{-1} = \begin{pmatrix}
\frac{x-S_N}{\gamma^{(N)}} & -\frac{\gamma^{(N+1)}}{\gamma^{(N)}} \cr
1 & 0
\end{pmatrix}.
\eeq
\ep

\bex[GUE]
Hermite polynomials satisfy
\beq
p^{(N+1)}(x) = x p^{(N)}(x) - N p^{(N-1)}(x)
\eeq
\beq
\implies \quad \gamma^{(N+1)} \psi^{(N+1)}(x) = x \psi^{(N)}(x) - \gamma^{(N)} \psi^{(N-1)}(x).
\eeq

\eex

\bd[Iterated shift operator]\label{def:Iterated.Shift.Operator}
Let
\beq
\mathcal R^{(N,k)}(x) := \mathcal R^{(N+k-1)}(x)\mathcal R^{(N+k-2)}(x)\cdots \mathcal R^{(N)}(x),
\eeq
which is a matrix, polynomial of $x$ of degree at most $k$.
\ed

It is such that
\beq
\Psi^{(N+k)}(x) = \mathcal R^{(N,k)}(x) \Psi^{(N)}(x).
\eeq

\textbf{ODE for the wave function}:

\bp[ODE for $\Psi$]
If $V'(x)\in\CC(x) $ is a rational function, then the orthogonal polynomials satisfy an ODE with rational coefficients, which in matrix form can be written
\beq
\frac{d}{dx} \Psi^{(N)}(x) = \mathcal D^{(N)}(x) \Psi^{(N)}(x),
\eeq
where
\beq
\mathcal D^{(N)}(x) =  \begin{pmatrix}
\frac{-V'(x)}{2} & 0 \cr
0 & \frac{V'(x)}{2}
\end{pmatrix}
+
\begin{pmatrix}
w_{N-1,N-1}(x) & w_{N-1,N}(x) \cr
w_{N,N-1}(x) & w_{N,N}(x)
\end{pmatrix}
\begin{pmatrix}
0 & \gamma^{(N)} \cr -\gamma^{(N)} & 0
\end{pmatrix},
\eeq
and
\beq
\begin{aligned}
w_{N,N'}(x)
&=&& \frac{1}{\sqrt{h^{(N)} h^{(N')}}} \int_\gamma e^{-V(x')}dx' \  p^{(N)}(x') \frac{V'(x)-V'(x')}{x-x'} p^{(N')}(x') \cr
&=&& \left(\frac{V'(x)-V'(Q)}{x-Q}\right)_{N,N'}, \cr
\end{aligned}
\eeq
with the matrix $Q$ defined in \autoref{eq:defQ1MM}.

We have
\beq\label{eq:TrDn1MM0}
\Tr \mathcal D^{(N)}(x) = 0.
\eeq
\ep

\proof{}See \cite{BertolaEynard},\cite[p.~155]{bookEynardKimuraRibault}.\eproof

\bex[GUE]
Hermite polynomials satisfy
\beq
\frac{d}{dx}p^{(N)}(x) = N p^{(N-1)}(x)  = x p^{(N)}(x) - p^{(N+1)}(x),\quad \frac{d}{dx}p^{(N+1)}(x) = (N+1) p^{(N)}(x).
\eeq
Therefore
\beq
\frac{d}{dx} \begin{pmatrix} p^{(N-1)}(x) \cr p^{(N)}(x) \end{pmatrix}
= \begin{pmatrix}
x & -1 \cr
N & 0
\end{pmatrix}
\begin{pmatrix} p^{(N-1)}(x) \cr p^{(N)}(x) \end{pmatrix},
\eeq
which implies
\beq
\frac{d}{dx} \Psi^{(N)}(x)
=\mathcal D^{(N)}(x) \Psi^{(N)}(x),\quad
\mathcal D^{(N)}(x)= \begin{pmatrix}
-\frac12 x & \gamma^{(N)} \cr
-\gamma^{(N)} & \frac12 x
\end{pmatrix}.
\eeq
Notice that
\beq
\Tr \mathcal D^{(N)}(x) = 0,
\eeq
\beq
\det \left(y-\mathcal D^{(N)}(x)\right) = y^2  - \frac14 x^2 +(\gamma^{(N)})^2 = y^2-\frac14 x^2+N.
\eeq
The eigenvalues of $\mathcal D^{(N)}(x)$  are $y=\pm \frac12 \sqrt{x^2-4N}$.

\eex



\bp[Some properties of the shift operator]
We have:
\beq
\mathcal R^{(N)}(x) = \mathcal R^{(N)} (\text{Id} + x \mathcal R^{(N)'}),\quad \mathcal R^{(N)}(x)^{-1} = (\text{Id}- x \mathcal R^{(N)'}) (\mathcal R^{(N)})^{-1},
\eeq
with $\mathcal R^{(N)}=\mathcal R^{(N)}(0)$. And $(\mathcal R^{(N)'})^2=0$ nilpotent, where
\beq
\mathcal R^{(N)'}  = \begin{pmatrix}
0 & \frac{-1}{\gamma^{(N)}} \cr 0 & 0
\end{pmatrix}.
\eeq
\ep

\proof{}
Both properties follow from direct computation.
\eproof
\\This implies
\beq
\mathcal R^{(N)}(x_2)\mathcal R^{(N)}(x_1)^{-1} = \mathcal R^{(N)}(\text{Id}+(x_2-x_1) \mathcal R^{(N)'}) (\mathcal R^{(N)})^{-1}.
\eeq



\bp[Recursion and ODE for the adjoint section]
We have:
\beq
M^{(N+1)}(x) = \mathcal R^{(N)}(x) M^{(N)}(x) \mathcal R^{(N)}(x)^{-1}, \quad
\frac{d}{dx} M^{(N)}(x) = [\mathcal D^{(N)}(x), M^{(N)}(x)].
\eeq

\ep
\proof{}
Special case of \autoref{prop:Rec.and.ODE.adjoint}
\eproof

\textbf{Recursion and ODE for \texorpdfstring{$W_1^{(N)}(x)$}{W1}}:

We will now merely apply the formalism of \autoref{sec:genRec}
 and \autoref{sec:genODE}.

Notice that we have
\beq
W_1^{(N+k)}(x)
= \Tr \mathcal D^{(N)}_k(x)  M^{(N)}(x)+\frac{V'(x)}{2},
\eeq
where
\beq
\mathcal D^{(N)}_k(x)
:= \mathcal R^{(N,k)}(x)^{-1} \mathcal D^{(N+k)}(x) \mathcal R^{(N,k)}(x).
\eeq

\bt[Recursion]
At most 3 of the matrices $\mathcal D^{(N)}_0(x),\dots,\mathcal D^{(N)}_3(x)$ can be linearly independent (use \autoref{eq:TrDn1MM0}), therefore there exist some coefficients $C^{(N,k)}(x)$, polynomials of $x$, such that
\beq
\sum_{k=0}^3 C^{(N,k)}(x) \mathcal D^{(N)}_k(x)=-2x.
\eeq
In analogy with \autoref{thm:ODEforW1}, this implies
\beq \label{eq:linear.recursion.relation.W1(N)}
\sum_{k=0}^3 C^{(N,k)}(x) W_1^{(N+k)}(x) = -2x,
\eeq
i.e. $W_1^{(N)}(x)$ satisfies a linear recursion relation of order 3, with coefficients polynomials of $x$.

\et

\bex[Recursion of $W_1^{(N)}$ in GUE]
In the case of GUE $W_1^{(N)}$ satisfies \autoref{eq:linear.recursion.relation.W1(N)} with
\begin{equation}
\begin{aligned}
C^{(N,0)} = -N-1 \,, \quad
C^{(N,1)} =  x^2 - N - 3 \,, \quad
C^{(N,2)} = -x^2 + N \,, \quad
C^{(N,3)} =  N+2 \,.
\end{aligned}
\end{equation}
This gives
\beq\label{eq:recW1GUE1MM}
(N+2)W_1^{(N+3)}(x) - (x^2-N) W_1^{(N+2)}(x)  + (x^2-N-3) W_1^{(N+1)}(x)
-(N+1)W_1^{(N)}(x) =-2x.
\eeq

\eex

\bt[ODE]
Let
\beq
{\hat{\mathcal D}}^{(N,0)}(x)=\mathcal D^{(N)}(x),
\quad 
{\hat{\mathcal D}}^{(N,k+1)}(x)=\frac{d}{dx} {\hat{\mathcal D}}^{(N,k)}(x) + [{\hat{\mathcal D}}^{(N,k)}(x),\mathcal D^{(N)}(x)].
\eeq
At most 3 of the matrices ${\hat{\mathcal D}}^{(N,0)}(x),\dots,{\hat{\mathcal D}}^{(N,3)}(x)$ can be linearly independent, therefore there exist some coefficients ${\hat C}^{(N,k)}(x)$, polynomials of $x$, such that
\beq
\sum_{k=0}^3 {\hat C}^{(N,k)}(x) {\hat{\mathcal D}}^{(N,k)}(x)=2N.
\eeq
This implies
\beq
\sum_{k=0}^3 {\hat C}^{(N,k)}(x) \frac{d^k}{dx^k} \ W_1^{(N)}(x) =2N,
\eeq
$W_1^{(N)}$ satisfies an ODE of order 3 with polynomial coefficients.

\et

\bex[GUE]

$W_1^{(N)}(x)$ satisfies an ODE of order 3:
\beq\label{eq:ODEGUE1MM}
\left( x  + (4N-x^2) \frac{d}{dx} +\frac{d^3}{dx^3} \right)W_1^{(N)}(x) = 2N.
\eeq

\eex

\bex[Consequence for moments of GUE]\label{ex:GUE1MM}

Let the moments of traces of powers:
\beq
t_k^{(N)} = \int_{\gamma^N} d\mu(\Lambda) \Tr \Lambda^k \  \  = \int_\gamma x^k \rho_1^{(N)}(x)dx.
\eeq
We have $t_0^{(N)}=N$, and it is clear that for the Gaussian measure, which is symmetric under $\Lambda\to -\Lambda$, all the odd moments vanish $t^{(N)}_{2k+1}=0$.
The resolvent can be written
\beq
W^{(N)}_1(x) = \sum_{k\geq 0} t^{(N)}_k x^{-k-1}.
\eeq
The ODE \eqref{eq:ODEGUE1MM} implies a recursion relation for the even moments $t_{2k}$.
\begin{equation}
t^{(N)}_{2k} =
2 N \frac{2k-1}{k+1} t^{(N)}_{2k-2}
+
\frac{1}{2} \frac{(2k-3)(2k-2)(2k-1)}{k+1} t^{(N)}_{2k-4}\,, \quad k \geq 2,
\end{equation}
with initial terms $t^{(N)}_0 = N, t^{(N)}_2 = N^2$.

The recursion \eqref{eq:recW1GUE1MM} implies:
\beq
t^{(N+2)}_{k+2}-t^{(N+1)}_{k+2}
= (N+2)t^{(N+3)}_{k}+Nt^{(N+2)}_{k}-(N+3)t^{(N+1)}_{k}-(N+1)t^{(N)}_{k}.
\eeq

\eex

\textbf{Recursion and ODE for \texorpdfstring{$W_2^{(N)}$}{W2}}:

We have
\beq
\begin{aligned}
W_2^{(N+k)}(x_1,x_2)
&= \frac{1}{(x_1-x_2)^2} \Tr  && M^{(N)}(x_1) \mathcal R^{(N,k)}(x_1)^{-1}\mathcal R^{(N,k)}(x_2) \cr
& &&\qquad M^{(N)}(x_2) \mathcal R^{(N,k)}(x_2)^{-1}\mathcal R^{(N,k)}(x_1).
\end{aligned}
\eeq

At most 4 of the $2\times 2$ matrices $\mathcal R^{(N,k)}(x_1)^{-1}\mathcal R^{(N,k)}(x_2) M^{(N)}(x_2) \mathcal R^{(N,k)}(x_2)^{-1}\mathcal R^{(N,k)}(x_1)$ for $k=0,1,2,3,4$ can be linearly independent. By \autoref{def:Iterated.Shift.Operator} $\mathcal R^{(N,k)}(x)$ and its inverse are polynomial in $x$. Therefore,

\bp
There exist some coefficients $C_2^{(N,k)}(x_1;x_2)$, polynomials of $x_1$, such that
\beq
\sum_{k=0}^4 \alpha_2^{(N,k)}(x_1,x_2)  \mathcal R^{(N,k)}(x_1)^{-1}\mathcal R^{(N,k)}(x_2) M^{(N)}(x_2) \mathcal R^{(N,k)}(x_2)^{-1}\mathcal R^{(N,k)}(x_1) = 0,
\eeq
\ep 
i.e. $W_2^{(N)}(x_1,x_2)$ satisfies a linear recursion relation of order 3, with coefficients polynomials of $x_1$ (possibly transcendental of $x_2$).

Moreover, we get

\bp There exist some coefficients $\alpha_2^{(N,k)}(x_1;x_2) \in \mathbb{C}[x_1,x_2]$, such that
\beq
\sum_{k=0}^{4} \alpha_2^{(N,k)}(x_1,x_2)  (W_2^{(N+k)}(x_1,x_2)-\frac{1}{(x_1-x_2)^2}) = 0.
\eeq
\ep 
\proof
We apply \autoref{thm:Rec.for.Wn} for $ \Lieg = \mathfrak{gl}(2, \mathbb{C})$ .
\eproof

\textbf{Recursion and ODE for \texorpdfstring{$W_n^{(N)}, n \geq 3$}{higher correlators}}:

We have
\beq
W_n^{(N+k)}(x_1,\dots,x_n) =
\sum_{\sigma\in \mathfrak S_n^{1-cycle}} \hspace{-1pt}
(-1)^{\sigma} \frac{\Tr \prod_{i=1}^n \mathcal R^{(N,k)}(x_{\sigma^{i}(1)}) M^{(N)}(x_{\sigma^{i}(1)}) \mathcal R^{(N,k)}(x_{\sigma^{i}(1)})^{-1} }{\prod_{i=1}^n (x_{\sigma(i)}-x_i)}.
\eeq

We can write it:
\beq
W_n^{(N)}(x_1,\dots,x_n) = \Tr M^{(N)}(x_1) \ H_n^{(N)}(x_1;x_2,\dots,x_n),
\eeq
where $H_n^{(N)}(x_1;x_2,\dots,x_n)$ is a $2\times 2$ matrix, rational function of $x_1$.

\bt
At most $4$ of the $2\times 2$  matrices $\mathcal R^{(N,k)}(x_1)^{-1}H_n^{(N+k)}(x_1;x_2,\dots,x_n)\mathcal R^{(N,k)}(x_1)$ for $k=0,1,2,\dots,4$ can be linearly independent, therefore there exist some coefficients $C_n^{(N,k)}(x_1;x_2,\dots,x_n)$, polynomials of $x_1$, such that
\beq
\sum_{k=0}^4 C_n^{(N,k)}(x_1;x_2,\dots,x_n)  W_n^{(N+k)}(x_1,x_2,\dots,x_n) = 0.
\eeq

\et

Following \autoref{thm:Rec.for.Wn}, we can also write:
\beq
W_n^{(N+k)}(x_1,\dots,x_n) = \mathcal H^{(N+k)}_n(M^{(N)}(x_1),M^{(N)}(x_2),\dots,M^{(N)}(x_n)),
\eeq
where $\mathcal H^{(N)}_n(x_1;x_2,\dots,x_n)$ is a linear form on $\mathfrak {gl}(2,\CC)^{\otimes n}$, it belongs to the dual ${\mathfrak {gl}(2,\CC)^{\otimes n}}^*$, which has dimension $4^n$, therefore at most $4^n$ of $\mathcal H^{(N+k)}_n(x_1;x_2,\dots,x_n)$ for $k=0,1,\dots,4^n$ can be independent.

This implies

\bt
There exist some coefficients $\td C_n^{(N,k)}(x_1,x_2,\dots,x_n)\in \CC[x_1,x_2,\dots,x_n] $, such that
\beq
\sum_{k=0}^{4^n} \td C_n^{(N,k)}(x_1,x_2,\dots,x_n)  W_n^{(N+k)}(x_1,x_2,\dots,x_n) = 0.
\eeq

\et

Similarly, we get an ODE for $W_n^{(N)}(x_1, \ldots, x_n)$ as shown in \autoref{sec:ODE.higher.correlators}. We begin with
\beq
\frac{d^k}{dx^k_1} W_n^{(N)}(x_1,\dots,x_n)
=
\Tr M^{(N)}(x_1) \ H_{k,n}^{(N)}(x_1;x_2,\dots,x_n),
\eeq
where $ H_{k,n}^{(N)}(x_1;x_2,\dots,x_n)$ is computed recursively
\beq
\begin{aligned}
H_{0,n}^{(N)} (x_1;x_2,\dots,x_n)
&=
H_{n}^{(N)} (x_1;x_2,\dots,x_n)
\\
H_{k+1,n}^{(N)}(x_1;x_2,\dots,x_n)
&=
[H_{k,n}^{(N)}(x_1;x_2,\dots,x_n),\mathcal D^{(N)}(x_1)]
\\
& \hspace{15mm}
+
\frac{d}{dx_1}H_{k,n}^{(N)}(x_1;x_2,\dots,x_n).
\end{aligned}
\eeq
Hence,
\bt
There exist some coefficients ${\hat C}^{(N,k)}_n(x_1;x_2,\dots,x_n)$, polynomials of $x_1$, such that
\beq
\sum_{k=0}^4 {\hat C}^{(N,k)}_n(x_1;x_2,\dots,x_n) \frac{d^k}{dx_1^k} W_n(x_1,\dots,x_n)=0.
\eeq
\et

\bt
There exist some coefficients ${\td C}^{(N,k)}_n(x_1,\dots,x_n) \in \CC[x_1,\dots,x_n]$, such that
\beq
\sum_{k=0}^{4^n} {\td C}^{(N,k)}_n(x_1,\dots,x_n) \frac{d^k}{dx_1^k} W_n(x_1,\dots,x_n)=0.
\eeq
\et

\textbf{Spectral density and correlations}:

As we mentionned in \autoref{cor:sameODEWrho}, the densities $\rho_n$ obey the same linear relations (ODE and recursion) as resolvents $W_n$.

\subsection{2-Matrix model}
\label{sec:2mm}
All the steps mentioned so far can be applied to the 2-matrix model.
The only difference with the 1-matrix model is that the Lie algebra is not $\mathfrak sl(2)$ but $\mathfrak sl(\td d)$ for some $\td d\geq 2$, i.e. matrices can have higher dimensions than $2$.

The link between the 2-matrix model and the formalism needed here, i.e. the wave functions $\Psi(x)$, the connection $\nabla = d-\mathcal D(x)dx$, the adjoint sections $M(x.E)$ and the determinantal correlators $W_n(x_1.E_1,\dots,x_n.E_n)$ can all be found in the literature \cite{BertolaEynard,BergereEynard,bookEynardKimuraRibault,MehtaBook}, let us recall them in short.

The following  defines what is called the ``2-matrix model" measure, actually its marginal measure for eigenvalues.

\bd[Matrix-eigenvalues measure]
Let two potentials $V$ and $\td V$ be such that $V'\in \CC(x)$, and $\td V'\in \CC(x)$.
Let $\Lambda=\diag(\Lambda_1,\dots,\Lambda_N)$ and $\td\Lambda=\diag(\td\Lambda_1,\dots,\td\Lambda_N)$ be two $N\times N$ diagonal matrices. We define
\beq
Z^{(N)} = \int_{\gamma^N\times \td\gamma^N}  d\Lambda d\td\Lambda \  \Delta(\Lambda) \Delta(\td\Lambda) \ \det\left(e^{\Lambda_i \td\Lambda_j}\right)e^{-\Tr \left( V(\Lambda)+\td V(\td \Lambda) \right)},
\eeq
and we define the marginal measure for $\Lambda\in \gamma^N $:
\beq
d\mu^{(N)}\Lambda = \frac{1}{Z^{(N)}} d\Lambda \int_{\td\gamma^N}  d\td\Lambda \  \Delta(\Lambda) \Delta(\td\Lambda) \ \det\left(e^{\Lambda_i \td\Lambda_j}\right)e^{-\Tr \left( V(\Lambda)+\td V(\td \Lambda) \right)},
\eeq
and the marginal measure for $\td\Lambda\in \td\gamma^N $:
\beq
d\td\mu^{(N)}\td\Lambda = \frac{1}{Z^{(N)}} d\td\Lambda \int_{\gamma^N}  d\Lambda \  \Delta(\Lambda) \Delta(\td\Lambda) \ \det\left(e^{\Lambda_i \td\Lambda_j}\right)e^{-\Tr \left( V(\Lambda)+\td V(\td \Lambda) \right)}.
\eeq
We denote
\beq
d = 1+\deg V', \quad
\td d = 1+\deg \td V',
\eeq
where $\deg V'$ is the sum of degrees of all poles, including the pole at $\infty$.
(if $V$, resp. $\td V$ is a polynomial, then $d=\deg V$, resp. $\td d =\deg \td V$).

\ed

\bd[Wave functions]
Let
\beq
p^{(N)}(x) := \int_{\gamma^N}  d\mu^{(N)}\Lambda \ \ \det(x-\Lambda), \quad
q^{(N)}(y) := \int_{\td\gamma^N}  d\td\mu^{(N)}\td\Lambda \ \ \det(y-\td\Lambda).
\eeq
Let
\beq
h^{(N)} := \int_{\gamma\times \td\gamma}dx dy \,  p^{(N)}(x) q^{(N)}(y) e^{-(V(x)+\td V(y)-xy)}, \quad
\gamma^{(N)} = \sqrt{h^{(N)}/h^{(N-1)}}.
\eeq
It is well known \cite{MehtaBook,BergereEynard} that $p^{(N)}, q^{(N)}$ form a biorthogonal family:
\beq
\int_{\gamma\times \td\gamma} dx dy \, p^{(N)}(x) q^{(\td N)}(y) \ e^{-(V(x)+\td V(y)-xy)}   = h^{(N)} \delta_{N,\td N}.
\eeq
We define:
\beq
\psi^{(N)}(x) := \frac{e^{-V(x)}}{\sqrt{h^{(N)}}} \ p^{(N)}(x)
\quad , \quad
\td\psi^{(N)}(y) := \frac{e^{-\td V(y)}}{\sqrt{h^{(N)}}} \ q^{(N)}(y),
\eeq
and we define the vector of dimension $\td d$:
\beq
\vec\psi^{(N)}(x) = \begin{pmatrix}
\psi_j^{(N-\td d+1)}(x) \cr \vdots \cr  \psi_j^{(N)}(x)
\end{pmatrix}.
\eeq

\ed

\bd[Dual wave functions]

We define the Hilbert transform
\beq\label{defphi2MM}
\phi^{(N)}(x) := \frac{e^{V(x)}}{\sqrt{h^{(N)}}}  \int_{\gamma\times \td\gamma } dx'dy'\,  \frac{1}{x-x'} q^{(N)}(y') \ e^{-(V(x')+\td V(y')-x' y')},
\eeq
and the vector
\beq
\vec \phi^{(N)}(x) = \begin{pmatrix}
\phi^{(N-1)}(x) \cr \phi^{(N)}(x) \cr \vdots \cr \phi^{(N+\td d-2)}(x)
\end{pmatrix}.
\eeq
We also define the Fourier transforms,
\beq
\phi^{(N,i)}(x) := \int_{\td\gamma_i }dy' \phi^{(N)}(y') \ e^{x' y'},
\eeq
where $\td\gamma_i, \ i=1,\dots,\td d-1 $ form a basis of homology paths on which the integral can be defined.

We define the $\td d\times \td d$  matrix by its matrix elements $i,j=1,\dots,\td d$:
\beq\label{def:PhiN2MM}
(\Phi^{(N)}(x))_{j,i} =  \phi^{(N-\td d+i,j-1)}(x),
\eeq
where we defined $\phi^{(N,0)}(x)=\phi^{(N)}(x)$.

\ed

\subsubsection{Recursion and ODE for wave functions}


\bp[Recursion relation]
The wave function satisfies a recursion relation of order $\td d$, written:
\beq
x \psi^{(N)}(x) = \gamma^{(N+1)} \psi^{(N+1)}(x) + \sum_{k=0}^{\td d-1} Q_{N,N-k} \psi^{(N-k)}(x),
\eeq
or written in matrix form
\beq
\vec\psi^{(N+1)}(x) = \mathcal R^{(N)}(x) \vec\psi^{(N)}(x),\quad 
\mathcal R^{(N)}(x)
=
\begin{pmatrix}
0 & 1 & &  \cr
0 &  & \ddots  & \cr
0 &   & & 1 \cr
\frac{-Q_{N,N-\td d+1}}{\gamma^{(N+1)}} &  \dots &\frac{-Q_{N,N-1}}{\gamma^{(N+1)}} & \frac{x-Q_{N,N}}{\gamma^{(N+1)}} \cr
\end{pmatrix}.
\eeq
$\mathcal R^{(N)}(x)$ is a matrix of size $\td d\times \td d$, polynomial of $x$ of degree 1.
Its inverse is also polynomial of degree 1.

Similarly for the dual wave functions we have, for $N\geq 1$
\beq
x \phi^{(N)}(x) = \gamma^{(N)} \phi^{(N-1)}(x) + \sum_{k=0}^{\td d-1} Q_{N+k,N} \phi^{(N+k)}(x),
\eeq
which can be written
\beq
\vec\phi^{(N+1)}(x) = {\td{\mathcal R}}^{(N)}(x) \vec\phi^{(N)}(x),\quad
{\td{\mathcal R}}^{(N)}(x)
=
\begin{pmatrix}
\frac{x-Q_{N,N}}{\gamma^{(N)}} & \dots & \frac{-Q_{N+\td d-2,N}}{\gamma^{(N)}} &  \frac{-Q_{N+\td d-1,N}}{\gamma^{(N)}} \cr
1 &  &  &  0 \cr
  & \ddots  & & \vdots \cr
 &   & 1 & 0\cr
\end{pmatrix}^{-1}.
\eeq
\ep
\proof{}
See \cite[p.~6]{BertolaEynard}.
\eproof

\bd[Iterated shift operator]
Let
\beq
\mathcal R^{(N,k)}(x) := \mathcal R^{(N+k-1)}(x)\mathcal R^{(N+k-2)}(x)\dots \mathcal R^{(N)}(x),
\eeq
which is a matrix, polynomial of $x$ of degree at most $k$.
It is such that
\beq
\vec\psi^{(N+k)}(x) = \mathcal R^{(N,k)}(x) \vec\psi^{(N)}(x).
\eeq
Similarly
\beq
{\td{\mathcal R}}^{(N,k)}(x) := {\td{\mathcal R}}^{(N+k-1)}(x){\td{\mathcal R}}^{(N+k-2)}(x)\cdots {\td{\mathcal R}}^{(N)}(x),
\eeq
which is a matrix, polynomial of $x$ of degree at most $k$.
It is such that
\beq
\vec\phi^{(N+k)}(x) = {\td{\mathcal R}}^{(N,k)}(x) \vec\phi^{(N)}(x).
\eeq

\ed


The wave functions satisfy a differential system of order $\tilde d$ (see \cite{BertolaEynard}):
\bp[ODE]
We have:
\beq
\frac{d}{dx} \vec\psi^{(N)}(x) = \mathcal D^{(N)}(x) \vec\psi^{(N)}(x),
\eeq
where $\mathcal D^{(N)}(x)$ is a matrix of size $\td d\times \td d$, polynomial of $x$ of degree less or equal to degree of $V'$.
The matrix $\mathcal D^{(N)}(x)$ is given explicitely in \cite{BertolaEynard, BergereEynard}.

Similarly, there is a $\td d\times \td d$, matrix ${\td{\mathcal D}}^{(N)}(x)$ polynomial of $x$ of degree less or equal to degree of $V'$ such that
\beq
\frac{d}{dx} \vec\phi^{(N)}(x) = {\td{\mathcal D}}^{(N)}(x) \vec\phi^{(N)}(x).
\eeq

\ep

\subsubsection{Matrix Wave function}

In this short presentation, we bypass the definition of the full $\td d\times \td d$ matrix $\Psi^{(N)}(x)$, we shall admit the following propositions, and refer the reader to \cite{BergereEynard,bookEynardKimuraRibault} in case of interest.

\bp[Matrix wave function]
there exists a square invertible matrix of size $\td d \times \td d$
\beq
\Psi^{(N)}(x) \in GL(\td d,\CC),
\eeq
whose first column is $(\psi^{(N-\td d+1)}(x),\dots,\psi^{(N)}(x))$, and that satisfies for $N$ large enough:
\beq
\frac{d}{dx} \Psi^{(N)}(x) = \mathcal D^{(N)}(x) \Psi^{(N)}(x),
\eeq
and
\beq
\Psi^{(N+1)}(x) = \mathcal R^{(N)}(x) \Psi^{(N)}(x).
\eeq
And there exists a square invertible matrix of size $\td d \times \td d$ (given in \autoref{def:PhiN2MM})
\beq
\Phi^{(N)}(x) \in GL(\td d,\CC),
\eeq
whose first column is $(\phi^{(N-1)}(x),\dots,\phi^{(N+\td d-2)}(x))$, and that satisfies for $N$ large enough:
\beq
\frac{d}{dx} \Phi^{(N)}(x) = {\td{\mathcal D}}^{(N)}(x) \Phi^{(N)}(x),
\eeq
and
\beq
\Phi^{(N+1)}(x) = {\td {\mathcal R}}^{(N)}(x) \Phi^{(N)}(x).
\eeq
\ep

\proof{Proved in \cite{BertolaEynard, BergereEynard}.}
\eproof
\bt[Duality]
\beq
\Phi^{(N)^t}(x) \mathbb A^{(N)} \Psi^{(N)}(x) = \text{Id},
\eeq
where $\mathbb A^{(N)}$ is the $\td d\times \td d$ matrix
\beq
\mathbb A^{(N)}_{i,j} = A_{N-2+i,N-2+j},
\eeq
where
\beq
A_{N-1+i,N-j}
= \left\{ \begin{array}{l}
 Q_{N-1+i,N-j}   \quad \text{ if } 0<i+j < \td d \cr
 - \gamma^{(N-1)}    \quad \text{ if } i=j=0 \cr
0    \quad \text{ otherwise }
\end{array}\right.
\eeq
This implies
\beq
\Psi^{(N)^{-1}}(x) = \Phi^{(N)^t}(x) \mathbb A^{(N)},\quad
\mathbb A^{(N)} \mathcal D^{(N)}(x) = - {\td{\mathcal D}}^{(N)}(x) \mathbb A^{(N)}.
\eeq

\et

\proof{}Proved in \cite{BertolaEynard, BergereEynard}.
\eproof

\subsubsection{Adjoint section}

\bd
We define:
\beq
M^{(N)}(x) = \Psi^{(N)}(x) E \Psi^{(N)}(x)^{-1}, \quad
E=\diag(1,0,\dots,0),
\eeq
or equivalently
\beq
M^{(N)}(x) = \Psi^{(N)}(x) E \Phi^{(N)}(x)^{t} \mathbb A^{(N)}.
\eeq

\ed

\bp
We have:
\beq
\left(M^{(N)}(x)\right)_{i,j} = \psi^{(N-\td d+i)}(x) \sum_{k=0}^{\td d-1}  \phi^{(N-2+k)}(x) A_{N-2+k,N-\td d+j},
\eeq
where
\beq
A_{N-1+i,N-j}
= \Biggl\{ \begin{array}{l}
 Q_{N-1+i,N-j}   \quad \text{ if } 0<i+j < \td d \cr
 - \gamma^{(N-1)}    \quad \text{ if } i=j=0 \cr
0    \quad \text{ otherwise }
\end{array}
\eeq
\ep
\proof{proved in \cite{BEHnlin.SI0108049}.} \eproof

\bt[Adjoint equations]The adjoint flat section $M^{(N)}$ satisfies:
\beq
\frac{d}{dx} M^{(N)}(x) = [ \mathcal D^{(N)}(x),M^{(N)}(x)],\quad
M^{(N+1)}(x) = \mathcal R^{(N)}(x) M^{(N)}(x)  \mathcal R^{(N)}(x)^{-1}.
\eeq
\et
\proof{Obvious.}\eproof

\subsubsection{Correlations}

\bd
Let the expectation of products of resolvents
\beq
\hat W^{(N)}_n(x_1,\dots,x_n) = \int_{\gamma^N} d\mu^{(N)}\Lambda \ \ \prod_{i=1}^n \Tr \frac{1}{x_i-\Lambda},
\eeq
and $W^{(N)}_n$ their cumulants.

Let also the densities
\beq
\hat \rho^{(N)}_n(x_1,\dots,x_n) = \int_{\gamma^N} d\mu^{(N)}\Lambda \ \ \prod_{i=1}^n \Tr \delta(x_i-\Lambda),
\eeq
and $\rho^{(N)}$ their cumulants.

\ed
\bp

The correlations of resolvents are the same as given by \autoref{def:Wngeneral}:
\beq
W^{(N)}_1(x.E) = \Tr \mathcal D^{(N)}(x) M^{(N)}(x.E),
\eeq
\beq
W^{(N)}_2(x_1.E_1,x_2.E_2) = \frac{1}{(x_1-x_2)^2} \Tr M^{(N)}(x_1.E_1)M^{(N)}(x_2.E_2),
\eeq
and for $n\geq 3$
\beq
W^{(N)}_n(x_1,\dots,x_n) = \sum_{\sigma\in \mathfrak S_n^{1-\text{cycle}}}
(-1)^\sigma \frac{\Tr \prod_{i=1}^n M^{(N)}(x_{\sigma^i(1)})}{\prod_{i=1}^n (x_{\sigma(i)}-x_i)}.
\eeq

\ep
\proof{See \cite{BergereEynard}}.\eproof

\subsubsection{Densities}

The same formulas as in \autoref{sec:1MMdensities} apply also in the 2-matrix model.

Notice from \autoref{defphi2MM} that $\phi^{(N)}(x)$ is discontinuous across $\gamma$:
\beq
\operatorname{Disc} \phi^{(N)}(x)  = \frac{1}{\sqrt{h^{(N)}}} \int_{\td \gamma} e^{-\td V(y')+xy'}q^{(N)}(y') dy'.
\eeq

This implies that resolvents $W_n$ are discontinuous, and due to the general Stieltjes theory, the density of eigenvalues and correlations, are the discontinuities of resolvents.
This gives the same formulas as in \autoref{sec:1MMdensities}.

\subsubsection{Linear ODE, Recursions}

As a consequence:

\bp
$W_1^{(N)}$ satisfy an ODE and a linear recursion of order $r$ with polynomial coefficients, where $r=\td d^2 = \dim \mathfrak gl(\td d)$.
\ep
\proof{
The ODE follows from \autoref{thm:ODEforW1} where we set $\mathfrak g = \mathfrak gl(\td d)$ and $\D(x) \rightarrow \D^{(N)}(x)$. The linear recursion is given by \autoref{thm:Rec.W1} with $\mathfrak g = \mathfrak gl(\td d)$.
}
\eproof
\bp
We have the following properties:
\begin{enumerate}
\item $W_n^{(N)}(x_1,\dots,x_n)$ satisfies an ODE of order $r$ with respect to the variable $x_1$ with  coefficients polynomials of $x_1$.

\item $W_n^{(N)}(x_1,\dots,x_n)$ satisfy an ODE with respect to the variable $x_1$  and a linear recursion  of order $r^n$ with coefficients polynomials of $(x_1,\dots,x_n)$.
\end{enumerate}
\ep
\proof{The first property follows from \autoref{thm:ODEforW_ncoef}.
For the second statement, the ODE follows from \autoref{thm:ODEforW_ncoefs} and the linear recursion follows from \autoref{thm:Rec.for.Wn}. 
}
\eproof

%
%
%
%
%
%
%
%
%
%
%
%
%
%

\subsection{Minimal models}
\label{sec:MM}

Our formalism is also well suited for minimal models.

Minimal models \cite{di1996conformal} arise as limits of matrix models.
They can also be defined independently of matrix models, from integrable hierarchies, KdV, KP, Toda, see \cite{BBT}.
They are classified by two integers $(p,q)$, in short, the $(p,q)$ minimal model can be formulated in terms of a differential system of order $q$, with polynomial coefficients whose degree is bounded by $p$.

\subsubsection{Minimal models \texorpdfstring{$q=2$}{q=2}}

This is the case of order 2 ODE, i.e. $\mathfrak sl(2,\CC)$.


Let $U(t)$ be a function of time $t$ that we shall precise below,
but let us already mention that it will be solution to some non-linear ODE having the Painlevé property.
In all what follows, there will be functions of $x$ and of time $t$.
We shall denote $f'=df/dx$ and $\dot f=df/dt$.

\bd[Gelfand-Dikii polynomials]
We define by recursion the following differential polynomials $R_n(U)\in \mathbb Q[U,\dot U, \ddot U, \dddot U,\dots]$
\beq
R_0(U)=2,
\eeq
\beq
\dot R_{n+1} = -2 U \dot R_n - \dot U R_n + \frac14 \dddot R_n,
\eeq
and we choose the primitive that is homogeneous (in $U$ and ${} \ \ddot{} \ {}$).

The first few terms are
\beq
R_0 =2, \quad
R_1 = -2U, \quad
R_2 = 3U^2 - \frac12 \ddot U, \quad
R_3 = -5 U^3 +\frac52 U \ddot U +\frac54 \dot U^2 - \frac18 \ddddot U.
\eeq
\ed

\bd[Lax pair]
Let
\beq
\mathcal R(x,t)
\coloneqq \begin{pmatrix}
    0 & 1 \cr
    x+2U(t) & 0
\end{pmatrix},
\eeq

define
\beq
B_n(x,U) \coloneqq  \frac12 \sum_{j=0}^n x^{n-j} R_j(U),
\eeq

and 
\beq
\mathcal D_n(x,t) \coloneqq  \begin{pmatrix}
    -\frac12 \dot B_n & B_n \cr
    (x+2U)B_n -\frac12 \ddot B_n & \frac12 \dot B_n
\end{pmatrix}.
\eeq
Notice that  $\mathcal R(x,t)$ and  $\mathcal D_n(x,t)$ are traceless, so that they belong to the Lie algebra $\mathfrak sl(2,\CC) $.

\ed

\bp[Gelfand Dikii]
\label{prop:GD}
We have
\beq
\dot{\mathcal D}_n(x,t)
+ [\mathcal D_n(x,t),\mathcal R(x,t)]
= - \dot R_{n+1}
\begin{pmatrix}
    0 & 0 \cr
    1 & 0
\end{pmatrix}.
\eeq
\ep

\proof{It is the classical Gelfand-Dikii theorem \cite{GD, BBT}.}\eproof

\bp[Compatibility and wave function]
Let $m\geq 0$ be an integer.
Let $\vec t=(t_0,t_1,t_2,\dots,t_m)$ be a set of parameters (called higher times).
Let
\beq
\mathcal D(x,t,\vec t) = \sum_{k=0}^m t_k \mathcal D_k(x,t).
\eeq

$\mathcal D(x,t;\vec t)$ satisfies the Lax equation
\beq
\frac{d}{dt}{\mathcal D}(x,t;\vec t)
+ [\mathcal D(x,t;\vec t),\mathcal R(x,t)] = \frac{d}{dx} \mathcal R(x,t),
\eeq
if and only if $U$ is solution to the equation
\beq\label{eq:EQforUq2}
\sum_{k=0}^m t_k R_{k+1}(U) = -t.
\eeq
In this case, there exists a matrix $\Psi(x,t;\vec t)$ such that
\beq
\frac{d}{dx} \Psi = \mathcal D \Psi , \quad
\frac{d}{dt} \Psi = \mathcal R \Psi.
\eeq
\ep
\proof{This is an immediate corollary of \autoref{prop:GD}.}\eproof

\bex[$m=0$ : Airy]

take $\vec t=(1)$, \autoref{eq:EQforUq2} reduces to:
\beq
-2U = -t.
\eeq
We have
\beq
\mathcal R(x,t) = \begin{pmatrix}
    0 & 1 \cr
    x+t & 0
\end{pmatrix}
=
\mathcal D(x,t).
\eeq
The matrix $\Psi(x,t)=\Psi(x+t)$ satisfies the Airy equation, and is worth:
\beq
\Psi(x,t)=\Psi(x+t) = \begin{pmatrix}
Ai(x+t) & Bi(x+t) \cr
Ai'(x+t) & Bi'(x+t)
\end{pmatrix}.
\eeq

\eex

\bex[$m=1$ : Painlevé 1]
\cite{Painleve}
take $\vec t=(0,1)$, then \autoref{eq:EQforUq2} is the Painlev\'e 1 equation:
\beq
3 U^2 - \frac12 \ddot U = -t,
\eeq
\beq
B_1(x,t) = x-U(t),
\eeq
and
\beq
\mathcal D_1(x,t) = \begin{pmatrix}
    \frac12 \dot U & x-U \cr
    (x+2U)(x-U) + \frac12 \ddot U & -\frac12 \dot U
\end{pmatrix}.
\eeq
Its characteristic polynomial is
\beq
\begin{aligned}
\det(y-\mathcal D_1(x,t))
&=&& y^2 - \frac14 \dot U^2 -(x+2U)(x-U)^2 - \frac12 \ddot U (x-U) \cr
&=&& y^2 - x^3 - (3 U^2 -\frac12 \ddot U)x -2U^3 - \frac14 \dot U^2 + \frac12 U \ddot U \cr
&=&& y^2 - x^3 +tx  -2U^3 - \frac14 \dot U^2 + \frac12 U \ddot U \cr.
\end{aligned}
\eeq
Remark that
\beq
\begin{aligned}
\frac{d}{dt}\left(-2U^3 - \frac14 \dot U^2 + \frac12 U \ddot U \right)
&=&& -6 U^2 \dot U - \frac12 \dot U \ddot U + \frac12 \dot U \ddot U + \frac12 U \dddot U \cr
&=&& U(-6 U \dot U  + \frac12  \dddot U) \cr
&=&& U \frac{d}{dt}(-3 U^2  + \frac12  \ddot U) \cr
&=&& U.
\end{aligned}
\eeq
\eex

\subsubsection{ODE with respect to time for \texorpdfstring{$W_1$}{W1}}

\bp[ODE for $W_1$]
$W_1$ satisfies the following ODE of order 3:
\beq
W_1 = \sum_{k=0}^m t_k \left(  (2(x+2U)B_k - \frac12 \ddot B_k) \dot W_1 +\frac12 \dot B_k \ddot W_1 - \frac12 B_k \dddot W_1 \right).
\eeq
\ep

\begin{proof}
Using the formula $W_1=\Tr(\mathcal{D}M)$ we have:
\beq
\begin{aligned}
    \frac{d}{dt}W_1&=&&\Tr(\frac{d}{dt}\mathcal{D}M+\mathcal{D}\frac{d}{dt}M)\cr
    \dot W_1&=&&\Tr((\mathcal{\dot D} +[\mathcal{D},\mathcal {R}])M)\cr
    \dot W_1&=&&\Tr(\mathcal{R}'M)\quad (\textit{By using the compatibility equation}).
\end{aligned}
\eeq
Therefore
\begin{equation}
    \frac{d^k}{dt^k}W_1=\Tr(\mathcal{D}_k M)=\Tr(\mathcal{\dot D}_{k-1}+[\mathcal{D}_{k-1},\mathcal{R}])M,\quad \mathcal{D}_0=\mathcal{D}=\sum_{k=0}^m t_k \mathcal{D}_k,
\end{equation}
we have

\begin{equation}
    \mathcal{D}_1=\mathcal{R}'=\begin{pmatrix}
    0&0\\
    1&0
    \end{pmatrix},\quad \mathcal{D}_2=\begin{pmatrix}
    -1&0\\
    0&1
    \end{pmatrix},\quad \mathcal{D}_3=\begin{pmatrix}
    0&-2\\
    2(x+2U)&0
    \end{pmatrix}.
\end{equation}
Also we remark that
\begin{equation}
    (x+2U)\mathcal{D}_1-\frac{1}{2}\mathcal{D}_3=\begin{pmatrix}
    0&1\\
    0&0
    \end{pmatrix},
\end{equation}
therefore
\begin{equation}
    \mathcal{D}_0=\sum_{k=0}^{m}t_k \bigg[B_k((x+2U)\mathcal{D}_1-\frac{1}{2}\mathcal{D}_3)+\frac{1}{2}\dot B_k \mathcal{D}_2+((x+2U)B_k-\frac{1}{2}\ddot B_k) \mathcal{D}_1\bigg].
\end{equation}

Hence
\begin{equation}
W_1=\sum_{k=0}^{m}t_k \bigg[(2(x+2U)B_k-\frac{1}{2}\ddot B_k)\dot W_1+\frac{1}{2}\dot B_k \dot W_1-\frac{1}{2}B_k\dddot W_1\bigg].
\end{equation}
\end{proof}
\eproof

\bex[Airy $m=0$, ODE $W_1$]

\begin{equation}
W_1 - 2(x+t) \dot{W}_1 + \frac{1}{2}  \dddot{W}_1 = 0.
\end{equation}

\eex

\bex[Painlevé 1, $m=1$, ODE $W_1$]
\begin{equation}
W_1 - (2x^2 + 2Ux-U^2 + t ) \dot{W}_1
+ \frac{1}{2} \dot{U} \ddot{W}_1 +\frac{1}{2} (x-U)  \dddot{W}_1 = 0.
\end{equation}
\eex

\subsubsection{Minimal models general case}
Minimal models can also be defined for rank $q\geq 2$ systems.
Let us give the example $(4,3)$ \cite{Berg15}, also called ``Ising model" (minimal models are related to conformal field theories and the $(4,3)$ model is the Ising conformal field theory \cite{BookYellowPagesCFT}) :

\beq
R(x,U) = \begin{pmatrix}
0 & 1 & 0 \cr
0 & 0 & 1 \cr
x + \frac{3}{2}\hbar \dot U & 3U & 0
\end{pmatrix},
\eeq

\begin{align}
\mathcal{D}(x,U) =
\begin{pmatrix}
2 U^2 - \frac{1}{6} \hbar^2 \ddot U
& x + \frac{1}{2} \hbar \dot U
& -U
\\
-Ux + \frac{5}{2} \hbar U \dot U - \frac{1}{6} \hbar^3 \dddot U
& -U^2 + \frac{1}{3} \hbar^2 \ddot U
&x - \frac{1}{2} \hbar \dot U
\\
x^2 + \hbar^2(\frac{7}{4} \dot{U}^2 + \frac{5}{2}U \ddot U) - \frac{1}{6} \hbar^4 \ddddot U
&
2Ux - \hbar U \dot U + \frac{1}{6} \hbar^3 \ddot U
&
-U^2 - \frac{1}{6} \hbar^2 \ddot U
\end{pmatrix}.
\end{align}

The Lax equation implies
\begin{align}
- \frac{1}{6} \hbar^4 \ddddot U + 3 \hbar^2 U \ddot U + \frac{3}{2} \hbar^2 \dot{U}^2 -4  U^3 =  t.
\end{align}

The ODE, with respect to time, satisfied by $W_1$ is
\begin{equation}
\sum_{k=0}^8 \alpha_k(x,t) \frac{d^k}{dt^k} W_1(x,t) = 0,
\end{equation}
where
\begin{equation}
\begin{aligned}
\alpha_0(x,t)
&=
9 \hbar^{2}  \dot{U},
\\
\alpha_1(x,t)
&=
- \frac{3 \hbar  }{4} \left(4 \hbar^{4} U {U}^{(5)} - 6 \hbar^{4} \dot{U} \ddddot{U} - 60 \hbar^{2} U^2 \dddot{U} \right.
\\
& \qquad \qquad
\left.
- 54 \hbar^{2} U \dot{U} \ddot{U} + 63 \hbar^{2} \dot{U}^{3}
+ 36 x^{2} \dot{U} + 72 U^3 \dot{U}\right),
\\
\alpha_2(x,t)
&=
- \frac{39 \hbar^{4} U \ddddot{U}}{2} + 15 \hbar^{4}  \dot{U} \dddot{U} + \frac{369 \hbar^{2}  U^2 \ddot{U}}{2}
\\
&
+ \frac{189 \hbar^{2}  U \dot{U}^{2}}{4} + 27 U x^{2} - 108  U^4,
\\
\alpha_3(x,t)
&=
- \frac{105 \hbar^{3} U \dddot{U}}{2} + \frac{75 \hbar^{3}  \dot{U} \ddot{U}}{2} + 189 \hbar  U^2 \dot{U},
\\
\alpha_4(x,t)
&=
- \frac{147 \hbar^{2} U \ddot{U}}{2} + 36 \hbar^{2}  \dot{U}^{2} + 81  U^3,
\\
\alpha_5(x,t)
&=
- 36 \hbar  U \dot{U},
\\
\alpha_6(x,t)
&=
- 18  U^2,
\\
\alpha_7(x,t)
&=
- \hbar  \dot{U},
\\
  \alpha_8(x,t) &= U.
\end{aligned}
\end{equation}

\subsection{Schlesinger systems}
\label{sec:Ss}
\subsubsection{Definition}

The Schlesinger system \cite{Schlesinger} with Lie algebra $\Lieg$ and $N$ points $z_1,\dots,z_N$, is defined as follows:
\beq
\mathcal D(x) = \sum_{i=1}^N \frac{A_i}{x-z_i},
\quad A_i\in \Lieg, \ \sum_{i=1}^N A_i=0.
\eeq
(remark: dropping the constraint $\sum_{i=1}^N A_i=0$ is equivalent to choosing a pole $z_{N+1}=\infty$ and take $A_\infty = -\sum_{i=1}^N A_i$).

We choose fixed monodromies:
\beq
\operatorname{eigenvalues}(A_i) = \diag(\theta_{i,1},\dots,\theta_{i,r}) = \theta_i \in \Lieh = \text{Cartan algebra}.
\eeq
and we fix the eigenvectors of $A_i$ by the Schlesinger equations:
\beq
\frac{d A_i}{dz_j} = [A_j,A_i].
\eeq

\subsubsection{Schlesinger \texorpdfstring{$\mathfrak sl(2)$, $N=3$}{sl(2), N=3}}

Take the points $0,1,\infty$ and
\beq
\mathcal D(x) = \frac{A_0}{x}+\frac{A_1}{x-1},
\eeq
with
\beq
A_0 = \begin{pmatrix}
\theta_0 & 0 \cr 0 & -\theta_0
\end{pmatrix},
\eeq
\beq
A_1= \frac{1}{2\theta_0} \begin{pmatrix}
\theta_\infty^2-\theta_1^2-\theta_0^2 & 2\theta_0\theta_1 +\theta_\infty^2-\theta_1^2-\theta_0^2  \cr 2\theta_0\theta_1 - (\theta_\infty^2-\theta_1^2-\theta_0^2) & -\theta_\infty^2+\theta_1^2+\theta_0^2
\end{pmatrix},
\eeq
\beq
A_\infty = -A_0-A_1= \frac{-1}{2\theta_0} \begin{pmatrix}
\theta_\infty^2-\theta_1^2+\theta_0^2 & 2\theta_0\theta_1 +\theta_\infty^2-\theta_1^2-\theta_0^2  \cr 2\theta_0\theta_1 - (\theta_\infty^2-\theta_1^2-\theta_0^2) & -\theta_\infty^2+\theta_1^2-\theta_0^2
\end{pmatrix}.
\eeq
They are such that
\beq
\operatorname{eigenvalues}(A_i) = \diag(\theta_i,-\theta_i).
\eeq

We find
\beq
\sum_{k=0}^3 \alpha_k(x) \frac{d^k}{dx^k} W_1(x) = 0,
\eeq

with coefficients:
\beq
\begin{aligned}
\alpha_0(x,t)
&=
6\theta_0^2x^2 - 10\theta_0^2x + 4\theta_0^2 - 6\theta_1^2x^2 + 2\theta_1 ^2x - 4\theta_{\infty}^2x^3
\\
& \quad + 6\theta_{\infty}^2x^2 - 2\theta_{\infty}^2x + 4x^3 - 6x^2 + 2x,
\\
\alpha_1(x,t)
&= 2 x(x-1) \  \left( 2\theta^2_0x-2\theta^2_0-2\theta^2_1x-2\theta_{\infty}^2 x^2+2\theta_{\infty}^2x+7x^2-7x+1\right),
\\
\alpha_2(x,t)
&= 4(2x-1) x^2 (1-x)^2,
\\
\alpha_3(x,t)
&= x^3(1-x)^3.
\end{aligned}
\eeq

\section{Conclusion}
\label{sec:conc}

We have shown how to exploit the ODE satisfied by the wave function (flat section of a Lie group bundle) or equivalently the adjoint ODE satisfied by $M$ (flat section of the adjoint Lie algebra bundle) to derive, in a systematic way, linear ODE and recursions with polynomial coefficients for every correlator $W_n$.

\section*{Acknowledgments}

We would like to thank A. Giacchetto for helpful discussions on this topic.
This paper is a result of the ERC-SyG project, Recursive and Exact New Quantum Theory (ReNewQuantum) which received funding from the European Research Council (ERC) under the European Union's Horizon 2020 research and innovation programme under grant agreement No 810573.
D. M. acknowledges support by Onassis Foundation under scholarship ID: F ZR 038/1-2021/2022.
\printbibliography



%
%

\end{document}